\documentclass[draft]{agujournal2019}
\usepackage{url}
\usepackage{amssymb}
\usepackage{amstext}
\usepackage{lmodern}
\usepackage{amsmath}
\usepackage{mathtools}
\usepackage{booktabs}
\usepackage{enumitem}
\usepackage{colortbl}
\usepackage{float}
\usepackage{lineno}
\usepackage[inline]{trackchanges}
\usepackage{xcolor}
\makeatletter
\providecommand\NAT@parse{}
\makeatother
\usepackage[hidelinks,final]{hyperref}
\usepackage{apacite}

\draftfalse

\journalname{Wiley}

\begin{document}

\title{Time Distribution of Heavy Rainfall in Brazil: Empirical Huff Curves from 290,164 Sub-Daily Storm Events}

\authors{Marcus Nobrega Gomes Junior\textsuperscript{1}, Andr\'e Sim\~oes Ballarin\textsuperscript{2}, Paulo Tarso Sanches de Oliveira\textsuperscript{3}\thanks{Corresponding author}}

\affiliation{1}{Stanford University - Department of Earth System Science (ORCID: 0000-0002-8250-8195)}
\affiliation{2}{University of S\~ao Paulo - Department of Hydraulics and Sanitation, S\~ao Carlos School of Engineering (ORCID: 0000-0001-6997-8662)}
\affiliation{3}{University of North Carolina at Charlotte - Department of Civil and Environmental Engineering (ORCID: 0000-0003-2806-0083)}

\correspondingauthor{Paulo Tarso Sanches de Oliveira}{polivei2@charlotte.edu}

\begin{keypoints}
    \item We derived the first national scale empirical rainfall timing curves for Brazil from over a thousand rain gauge stations
    \item Storms in Brazil concentrate rainfall in the first quarter of their duration far more often than the classic Illinois reference
    \item Using Brazilian curves instead of the Illinois reference raises modeled flood design peak discharge by about eight percent
\end{keypoints}

\begin{abstract}
Temporal rainfall distributions are widely used in design storm construction; however, many countries with limited sub-daily records, including Brazil, still rely on frameworks developed under different hydroclimatic conditions, introducing potential bias in hydrological design. Here, we develop the first national-scale empirical Huff curves for Brazil using sub-daily rainfall observations. We compiled data from $3{,}164$ stations and, after quality control, retained $290{,}164$ storm events from $1{,}045$ stations spanning 2010 to 2025. Empirical cumulative mass curves were fitted with $7^{\mathrm{th}}$-degree polynomials at station, biome, state, and municipality scales. Our findings show a dominance of first-quartile ($Q_1$; front-loaded) storm patterns, occurring at $94.4\%$ of stations nationally, increasing to $99.2\%$ in the Amazon and Cerrado biomes and decreasing to $88.7\%$ in the Atlantic Forest. The national $Q_1$ median curve closely matches the Huff (1967) reference ($\mathrm{MAE}=0.045$; $D_{\max}=0.097$), with narrow bootstrap uncertainty (mean width $=0.006$). $Q_1$ dominance is robust to inter-event time definition, with $84.7\%$ of stations showing consistent classification across $2$ to $12$~h thresholds. We attribute this prevalence to thermally driven deep convection, consistent with observed afternoon initiation peaks. A Soil Conservation Service Curve Number (SCS-CN) experiment across $579$ headwater catchments shows that Brazilian curves increase design peak discharge by a median of $8\%$ and up to $11\%$ in the Cerrado relative to the Illinois reference, indicating potential underestimation when using non-local distributions. Biome, state, and municipality scale parameters are provided as open data and through an interactive platform for local design-storm applications.
\end{abstract}

\section*{Plain Language Summary}
\noindent Engineers designing storm drains, culverts, and flood control structures need to know not just how much rain falls during a storm, but how that rain is distributed over time, since a storm that releases most of its rain in the first hour produces a much sharper flood peak than one that rains steadily throughout. Brazil has lacked its own data driven description of this rainfall timing pattern and has instead relied on a pattern developed in Illinois, USA, in the 1960s. We analyzed sub-daily rainfall records from over a thousand Brazilian monitoring stations spanning 2010 to 2025 and found that Brazilian storms load their rainfall toward the beginning far more consistently than the Illinois pattern assumes, a result consistent with Brazil's frequent afternoon thunderstorms. When we used the Brazilian pattern instead of the imported one in a simple flood model, the modeled flood peak increased by about eight percent on average, and by more in the most convective regions. Because the imported pattern systematically underestimates flood peaks, Brazilian engineers should adopt the locally derived curves we provide, which are freely available online.

\section{Introduction}
\label{sec:intro}

Rapid urbanization alters catchment hydrological response by increasing impervious surfaces, runoff volumes, and peak flows, while climate change is expected to intensify short-duration extreme precipitation events, together placing increasing pressure on urban drainage systems \cite{Walsh2005UrbanStreamSyndrome,Fletcher2013UrbanHydrology,Westra2014ShortDurationExtremeRainfall,ArnbjergNielsen2013UrbanDrainageReview}. Consequently, the capacity of storm sewers, detention basins, green infrastructure, and flood-retention facilities to safely convey, detain, and store excess runoff has become a critical component of urban flood resilience \cite{Fletcher2013UrbanHydrology,MailhotDuchesne2010UrbanDrainageDesign,ArnbjergNielsen2013UrbanDrainageReview}. The hydraulic design of these components requires translating a statistical description of extreme rainfall, commonly expressed as a design storm, into a time-varying rainfall rate that can be routed through the drainage network \cite{MailhotDuchesne2010UrbanDrainageDesign,ArnbjergNielsen2013UrbanDrainageReview}. The hydraulic design of each component requires translating a statistical description
of extreme rainfall (the design storm) into a time-varying rainfall
rate that can be routed through the drainage network. This
translation demands two pieces of information: the total rainfall
depth associated with a given duration and return period, encoded in
intensity–duration–frequency (IDF) curves; and the temporal distribution
of that depth within the storm, which determines how rapidly the
drainage system is loaded and therefore governs peak discharge, maximum
water levels, and overflow volumes.

The temporal distribution of rainfall, often characterized by a storm hyetograph or rainfall mass curve, plays a fundamental role in watershed response. Even when total rainfall depth and storm duration are identical, differences in the timing and concentration of rainfall intensity can lead to  different runoff generation, peak flows, and flood magnitudes. Front-loaded storms generate high initial abstraction losses followed by rapid overland
flow on a progressively saturating surface, yielding a sharp,
early hydrograph peak. Back-loaded or uniform storms allow longer
periods of infiltration before surface runoff initiates, shifting
the peak later and broadening the hydrograph. Retention basins sized
for one pattern may overflow under another. Sewer pipes dimensioned
for a particular peak-flow rate may be undersized if the actual storm
produces a steeper rising limb than assumed in design. In short,
the temporal pattern of rainfall is as consequential for infrastructure
design as its intensity.

In engineering practice, temporal distributions are standardised
through dimensionless mass curves that express cumulative storm depth
as a fraction of total depth against normalised storm time. Several
frameworks have been proposed (e.g., the SCS Type curves and the
Chicago hyetograph), but the approach proposed by \citeA{Huff1967},
the four-quartile scheme, in particularly the first quartile rainfall distribution has become the most widely adopted
international reference for urban drainage.
In that framework, storms are grouped according to the quarter of
normalised storm duration in which the greatest depth falls: Q1
if rainfall is most intense in the first 25/
in the second quarter, and so on, and empirical cumulative mass
curves are derived for each class from a long gauge record.

\citeA{Huff1967} developed the original curves from 261 heavy storms
(minimum total depth 12.7~mm, i.e., 0.5~in) recorded on a
dense 49-gauge network in Illinois, USA, over 1955 to 1966. The
resulting four-quartile distributions, together with the expanded
tabulations of \citeA{HuffAngel1992}, have since been adopted in
hydrological studies, rainfall disaggregation procedures, and
drainage-design codes worldwide \cite{Bonta2004, Bezak2018}.
This widespread adoption persists despite the fact that the curves
were calibrated in a temperate continental climate fundamentally
different from the tropical and subtropical regimes where they are
most often applied without local verification.

The sensitivity of the hydrological response to the choice of design
storm temporal distribution has been repeatedly and sharply quantified.
\citeA{Bezak2018} reported peak-flow differences of up to 132/%
between storm types in a central-European context, while
\citeA{Abreu2017} found that substituting a Q1 Huff pattern for a
uniform distribution changed peak flow by 46/
by 57/
These findings establish that the selection of an appropriate temporal
pattern is not a secondary methodological concern but a primary source
of uncertainty in flood-risk assessment.

This recognition has driven Huff-curve derivations globally,
consistently showing that the dominant quartile and curve shape are
strongly climate-dependent. \citeA{AzliRao2010} found Q2 dominance
in Peninsular Malaysia ($\sim$5,800 storms), contrasting with
\citeA{Huff1967}'s Q1/Q2 subequal result. In Europe,
\citeA{Dolsak2016} showed that Huff curves from Slovenia's three
climate zones were all dissimilar from each other and from Illinois,
Malaysia, and Brazil. \citeA{Liang2017} found that 84/
Guangzhou fall in Q1 or Q2, with peak rainfall at $33 \pm 5\%$ of
storm duration, a front-loaded regime consistent with tropical
convection. In Colombia, \citeA{Corrales2026} derived regionalized
Huff curves for 270 short-duration storms, identifying three dominant
spatial types: early-peak, intermediate, and uniform.

Beyond design-storm applications, the Huff framework has also been
extended to stochastic disaggregation \cite{Bonta2004} and to
modified limb-description and intensity-threshold refinements
\cite{Dunkerley2022}, confirming that, as sub-daily networks
expand, the framework continues to evolve.

Despite the recognized importance of Huff curves and their implications for urban drainage design, in Brazil, a continental-scale country encompassing a wide range of hydroclimatic regimes influenced by the Intertropical Convergence Zone (ITCZ), South American Monsoon System (SAMS), South Atlantic Convergence Zone (SACZ), and extratropical cold fronts \citeA{Marengo2012}, still lacks a
national-level Huff-curve study.  Existing studies are sparse and geographically limited. For instance, \citeA{DAEE1983} adopted the original \citeA{Huff1967} curves without regional calibration for the Alto Tietê macrodrainage plan, whereas \citeA{SantaCatarina2021} and \citeA{Abreu2017} derived or evaluated local Huff curves for southern Brazil, both indicating a predominance of first-quartile (Q1) storms. Although these studies provide valuable regional insights, they do not offer national coverage, a standardized storm-selection methodology consistent with \citeA{Huff1967}, or a systematic evaluation of how Brazilian rainfall temporal patterns compare with the widely adopted Illinois reference across distinct hydroclimatic regions.

In view of this knowledge gap and its implications for hydrologic design, this study pursues three objectives. First, we develop national-scale empirical Huff curves for Brazil using storm-selection criteria consistent with \citeA{Huff1967}, and provide fitted curves at station, biome, state, and municipality scales. Second, we quantify deviations between Brazilian empirical Huff curves and the Illinois reference using mean absolute error (MAE), maximum absolute deviation ($D_\mathrm{max}$), and uncertainty bounds of regional median curves. Third, we characterize dominant storm-quartile patterns across Brazil and examine their spatial, seasonal, and diurnal variability in relation to key climatic drivers of rainfall.

We use the telemetric hydrometeorological network of the Agência Nacional de Águas e Saneamento Básico (ANA), which comprises more than 3,000 stations equipped with sub-daily rainfall gauges, providing an unprecedented opportunity to characterize storm temporal structure across Brazil. We further assess the robustness of the results to the definition of the inter-event time (IETD), evaluate their engineering relevance using a controlled SCS-CN design-hydrograph experiment, and provide polynomial coefficients and design-storm products as open supplementary data, complemented by an interactive web application.

The paper is structured as follows. Section~\ref{sec:data} describes
the study area, the ANA telemetric network, and the quality-control
procedure. Section~\ref{sec:methods} details the event extraction,
curve derivation, polynomial fitting, uncertainty quantification, IETD
sensitivity analysis, design-hydrograph experiment, and
web-application development. Section~\ref{sec:results} presents the
national, biome-scale, spatial, robustness, and hydrological-impact
results. Section~\ref{sec:discussion} interprets the dominant
storm-quartile mechanism, compares the findings with prior studies,
and discusses implications for Brazilian design-storm practice.
Section~\ref{sec:conclusions} summarises the main findings.

\section{Study area and data}
\label{sec:data}

\subsection{Hydroclimatic setting}

\begin{figure}[H]
  \centering
  \includegraphics[width=\columnwidth]{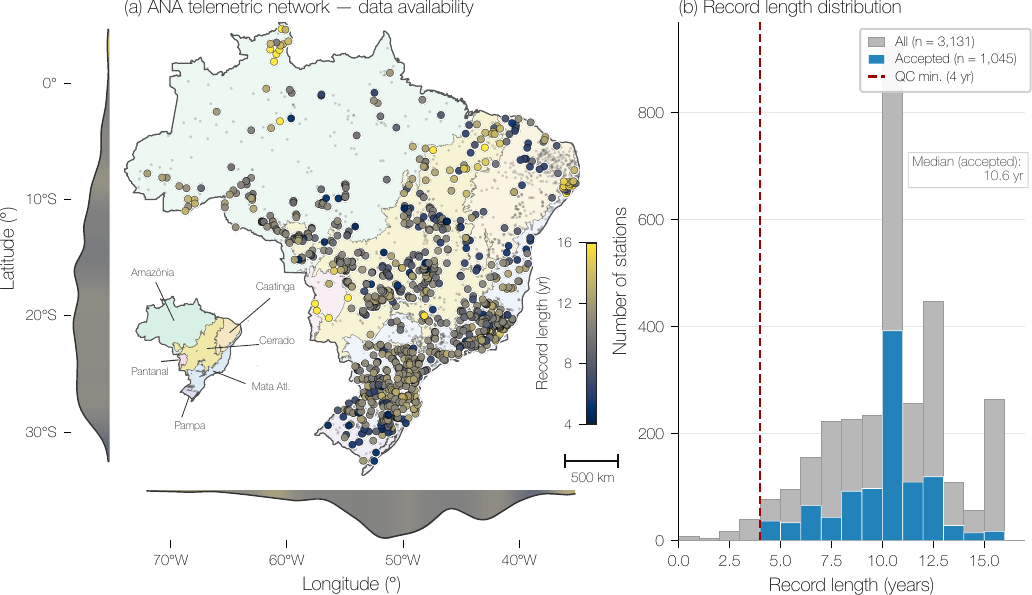}
  \caption{(a) Spatial distribution and record length of the 3{,}164 ANA
    telemetric stations across Brazil. Biome zones are shown with distinct
    colour fills (legend); state boundaries in dark grey. Station circles
    are coloured by years of available record (inset colour scale);
    light grey circles ($n=2{,}119$) indicate stations excluded from the
    analysis due to insufficient record length or excessive missing data.
    (b) Distribution of station record lengths for all catalogued stations
    (grey) and the 1{,}045 quality-control-accepted stations (colour-filled).
    Dashed red line: 4-year minimum quality-control threshold.
    Median record length for accepted stations: 10.6~years.}
  \label{fig:casestudy}
\end{figure}

Brazil spans latitudes 5°N–34°S and longitudes 35°W–74°W, a
continental territory of 8.5~million~km$^2$ encompassing six official
biomes defined by the Instituto Brasileiro de Geografia e Estatística
(IBGE): Amazônia, Caatinga, Cerrado, Mata Atlântica, Pampa, and
Pantanal (Fig.~\ref{fig:casestudy}). The country's precipitation
regime is governed by a complex interplay of large-scale atmospheric
circulation systems, orography, and ocean–land thermal contrasts
\cite{Marengo2012}.

The Amazônia biome, covering approximately 49\% of the national
territory, is characterised by an equatorial climate with year-round
deep convection driven by high sea-surface temperatures in the tropical
Atlantic, moisture recycling, and the seasonal migration of the
Intertropical Convergence Zone (ITCZ). Annual precipitation totals
typically exceed 2,000~mm, rising above 3,000~mm in the western
Amazon. The wet season (November–April) is associated with the
South American Monsoon System (SAMS) and the South Atlantic
Convergence Zone (SACZ), which together channel moisture from the
Amazon into southeastern Brazil. A distinct dry season occurs in
June–August in the southern and eastern sectors of the biome.
Despite its high annual total, much of the Amazon's rainfall is
produced by intense organised convective systems (including mesoscale
convective complexes) that generate some of the world's highest
lightning flash densities \cite{Zipser2006}.

The Caatinga, covering the semi-arid interior of the Northeast region,
presents one of the most challenging rainfall regimes in Brazil.
Total annual precipitation ranges from 300 to 800~mm, concentrated
in a brief wet season (February–April) driven by the southward
migration of the ITCZ. Events are rare but intense, often exceeding
50~mm~h$^{-1}$, and the long dry season imposes severe constraints
on the reliability of any sub-daily gauge record. The Cerrado, to
the south and west, is a tropical savannah with a strongly bimodal
annual cycle: a convective wet season from October to April, during
which the interior plateau receives 70–80\% of its annual rainfall
from afternoon mesoscale convective systems, followed by a near-total
dry season from May to September. This high seasonality and the
dominance of thermally forced convection make the Cerrado the biome
most likely to produce front-loaded storm temporal patterns.

The Mata Atlântica occupies the Atlantic coast from the far northeast
to Rio Grande do Sul state and encompasses the most complex storm
climatology in Brazil. Its topography, the Serra do Mar rising
abruptly from the coast to over 2,000~m within 50~km, and the
Serra da Mantiqueira reaching 2,800~m, generates intense
orographic enhancement during the wet season (October–March) and
during the passage of frontal systems from the south. Sea-breeze
circulations produce nocturnal rainfall maxima in coastal zones,
while inland valleys experience afternoon convection. Cold fronts
penetrating from the extratropics are the dominant forcing mechanism
in winter (June–August) and account for a significant fraction of
the annual total in the southern portion of the biome. This diversity
of storm-generating mechanisms, convective, orographic, and frontal,
produces a correspondingly diverse population of storm temporal
patterns, making the Mata Atlântica the biome most likely to deviate
from a single dominant quartile.

The Pampa (southern Brazil, Rio Grande do Sul) and Pantanal (western
Mato Grosso and Mato Grosso do Sul) represent the two smallest
biomes. The Pampa has a subtropical climate with no pronounced dry
season, high frequency of extratropical cyclone tracks from the
southwest, and frontal rainfall distributed relatively uniformly
through the year, the most Illinois-like climate in Brazil. The
Pantanal is an extensive tropical wetland with highly concentrated
summer convection (November–March, typically exceeding 1,200~mm)
and a pronounced dry season. Its flat topography and high soil-moisture
during the wet season allow deep convection to initiate readily,
producing intense and short-lived convective storms.

This diversity of climatic mechanisms, spanning equatorial deep
convection, semi-arid ITCZ-driven events, tropical savannah
afternoon convection, orographic and frontal rainfall, and
extratropical storm tracks, makes Brazil a uniquely challenging
and scientifically valuable test case for evaluating the
transferability of the \citeA{Huff1967} Illinois reference across
contrasting climatic regimes.

\subsection{ANA telemetric network and data acquisition}

Sub-daily rainfall data were downloaded from the ANA
\textit{DadosHidrometeorologicos} SOAP endpoint
(\url{https://telemetriaws1.ana.gov.br/ServiceANA.asmx}). The station
catalogue comprised 3,164 stations distributed across all 26 Brazilian
states and the Federal District. Data were retrieved for the period
1 January 2010 to 31 December 2025 in 90-day chunks to comply with
service limits.

The ANA \textit{Chuva} field is treated as interval rainfall depth
in millimetres, so that intensity is derived as $i = d / \Delta t$,
where $d$ is the recorded depth and $\Delta t$ is the timestep
inferred from the modal inter-record interval. The modal timestep was
15~min for 371 stations (35.5\% of accepted stations), 30~min for 133
(12.7\%), and 60~min for 541 (51.8\%).

\subsection{Quality control}

Each station time series was regularised onto a uniform time grid at
its inferred timestep, after which four quality filters were applied.
The first required a record span of at least four years, following
\citeA{BontaRao1988}, who demonstrated that this minimum is needed for
stable percentile estimation in event-based Huff analyses. The second
required the fraction of missing observations to remain at or below
20\%, consistent with the World Meteorological Organization guideline
that stations exceeding this threshold should be treated with caution
for climatological analysis \cite{WMO2018}. The third rejected any
station containing a near-complete calendar year (defined as one with
at least 95\% valid records) that nonetheless reported zero total
depth, since this pattern almost always indicates an instrument
failure rather than a genuinely dry year. Finally, interval depths
implying an intensity greater than 300~mm~h$^{-1}$ were set to missing
as physically implausible for tipping-bucket gauges.

Of the 3,164 stations, 1,045 (33.0\%) passed all filters
(Supplementary Table~S1). The dominant failure cause was excessive
missing data ($>$~20\%), affecting 93.1\% of the 2,086 rejected
stations. The spatial distribution of accepted stations and their
record lengths is shown in Fig.~\ref{fig:casestudy}.

\section{Methods}
\label{sec:methods}

\subsection{Event extraction}
\label{sec:event_extraction}

Rainfall events were identified by applying an inter-event time
definition (IETD) of 6~h to separate consecutive wet periods. This
value replicates the storm-separation criterion of \citeA{Huff1967}
and is consistent with recommendations for humid climates by
\citeA{RestrepoPosada1982} and \citeA{BontaRao1988}. The sensitivity
of results to this choice is evaluated in Section~\ref{sec:ietd}.

Within each extracted event, a minimum total depth of 12.7~mm
(0.5~in) and a minimum of four records were required, again following
\citeA{Huff1967}. Events exceeding 96~h in duration were discarded,
as such long windows typically encompass multiple distinct storm
systems rather than a single convective episode. Interval records
implying intensities above 300~mm~h$^{-1}$ were set to missing as
physically implausible for tipping-bucket gauges.

\subsection{Huff curve derivation}

For each qualifying event, the dimensionless cumulative mass curve
was computed as:
\begin{equation}
  F(\tau_i) = \frac{\sum_{j=1}^{i} d_j}{\sum_{j=1}^{n} d_j},
  \quad \tau_i = \frac{i}{n},
  \label{eq:cdf}
\end{equation}
where $d_j$ is the rainfall depth at the $j$-th record, $n$ is
the total number of records in the event, $\tau \in [0,1]$ is the
normalised storm time, and $F \in [0,1]$ is the normalised cumulative
depth. Quartile assignment followed \citeA{Huff1967}: the event was
classified as Q$k$ ($k = 1, 2, 3, 4$) according to the quarter of
$[0, 1]$ in which the greatest incremental depth fell:
\begin{equation}
  k = \underset{k \in \{1,2,3,4\}}{\arg\max}
        \left[ F\!\left(\tfrac{k}{4}\right) -
               F\!\left(\tfrac{k-1}{4}\right) \right].
\end{equation}

All individual-event mass curves were interpolated onto a common
$\tau$ grid with step 0.02 and the station median (50th-percentile)
curve was computed for each quartile. Percentile envelopes
(P10–P90 in 10-percentage-point steps) were also retained to
characterise intra-station storm variability.

\subsection{Polynomial fitting}
\label{sec:poly}

Median curves were represented by 7th-degree polynomials
($\hat{F}(\tau) = \sum_{m=0}^{7} a_m \tau^m$) following the
convention of \citeA{Huff1967}. Coefficients were estimated by
least-squares regression; polynomial values were subsequently
clipped to $[0, 1]$ and monotonised to enforce the CDF property.
The goodness of fit was uniformly $R^2 > 0.99$ at all spatial
scales.

\subsection{Goodness-of-fit metrics}
\label{sec:metrics}

Agreement between empirical and reference curves was quantified by
two metrics evaluated at $\tau = 0.1, 0.2, \ldots, 1.0$ on the
Huff (1967) reference grid:

\begin{align}
  \mathrm{MAE} &= \frac{1}{N} \sum_{i=1}^{N}
                  \bigl| \hat{F}(\tau_i) - F_\mathrm{ref}(\tau_i) \bigr|,
                  \label{eq:mae} \\
  D_\mathrm{max} &= \max_{i} \bigl| \hat{F}(\tau_i)
                   - F_\mathrm{ref}(\tau_i) \bigr|.
                   \label{eq:dmax}
\end{align}

MAE directly quantifies the average pointwise departure in units of
cumulative rainfall fraction. $D_\mathrm{max}$ captures the
worst-case departure at any storm phase, the quantity most critical
for design applications.

The Huff (1967) reference curves used for comparison were digitised
from the published median (50th-percentile) distributions and
represented as 7th-degree polynomials for computational evaluation.
Endpoints were constrained to exactly $\{0, 1\}$ after polynomial
evaluation.

\subsection{Regional aggregation}

Station median curves were aggregated to biome, state, and
municipality levels by computing the median across contributing
station curves at each $\tau$ point, then fitting a new 7th-degree
polynomial to the regional median. Municipality-level curves with
fewer than five contributing stations were flagged as unreliable.

\subsection{Bootstrap uncertainty quantification}
\label{sec:bootstrap}

Regional curve uncertainty was quantified by bootstrapping over
stations: $B = 2{,}000$ samples were drawn with replacement from the
set of station median curves within each region, and the national or
biome median curve was recomputed for each draw. The 2.5th and 97.5th
percentiles across draws form the 95\% confidence interval on the
regional median. Station-level uncertainty was approximated from the
inter-event percentile spread (P10–P90) using the standard-error-of-
the-median formula $\mathrm{SE} \approx 1.253\,\sigma/\sqrt{n}$,
where $\sigma$ is estimated from the P10–P90 range and $n$ is the
number of Q1 events at the station.

\subsection{Inter-event time definition (IETD) sensitivity analysis}
\label{sec:ietd_method}

To evaluate robustness to the IETD choice, the full pipeline was
re-executed at IETD $\in \{2, 4, 8, 12\}$~h in addition to the
baseline of 6~h. Four metrics were evaluated across all five runs:
(i) total qualifying event count and median event duration;
(ii) event-level quartile fraction;
(iii) station-level dominant-quartile agreement with the 6-h baseline;
(iv) MAE and $D_\mathrm{max}$ of the national Q1–Q4 median curves
relative to the Huff (1967) reference. All comparisons restrict the
station set to the 1,045 accepted stations to isolate the effect of
the IETD from changes in station membership.

\subsection{Design-hydrograph sensitivity experiment}
\label{sec:hydro_method}

To translate the difference between the original and updated curves into
an engineering-relevant quantity, we conducted a controlled
design-hydrograph experiment in which a single factor — the temporal
rainfall pattern — is varied while catchment, soil, terrain, design
depth, and storm duration are held fixed. The change in peak discharge
and in time-to-peak is therefore attributable solely to the hyetograph
shape. This experimental framework mirrors standard Brazilian engineering
practice, in which the SCS-CN method and the SCS dimensionless unit
hydrograph are the most widely adopted approaches for design-hydrograph
computation in small headwater catchments.

The experiment is restricted to small headwater catchments in which the
Soil Conservation Service curve-number (SCS-CN) method and the SCS
dimensionless unit hydrograph are applicable
\cite{NRCS2004, TR55_1986}. The SCS-CN method was developed for small,
infiltration-excess (Hortonian) agricultural watersheds and is invalid
where saturation-excess (variable-source-area) runoff dominates or where
rainfall cannot be treated as spatially uniform. We therefore selected
catchments from the HydroBASINS level-12 layer \cite{Lehner2013} using
explicit eligibility criteria: drainage area between 5 and 250~km$^2$
(small enough for a lumped unit hydrograph and uniform design rainfall);
and biome membership restricted to the Cerrado, Caatinga, Mata
Atlântica, and Pampa. The Amazônia and Pantanal biomes were excluded a
priori because their low-gradient, saturation-excess and wetland runoff
regimes violate the SCS-CN conceptual model. Of 30{,}866 eligible
catchments, a stratified random sample of 150 per biome (600 total) was
drawn.

For each catchment, the time of concentration was estimated using the \citeA{Kirpich1940} equation, based on main-channel length (derived from drainage area through Hack’s law) and channel slope. The latter was computed as the ratio between catchment relief (defined as the 5th–95th percentile elevation range) and channel length, using the Copernicus GLO-90 digital elevation model \cite{CopernicusDEM}. The design rainfall depth was taken from the gridded
Sherman-type intensity–duration–frequency (IDF) surfaces of
\citeA{GomesJuniorIDF}, derived from the Brazilian Daily Weather
Gridded Data (BR-DWGD; \citeNP{Xavier2022}) and satellite records.
The IDF model is $i = K\,T^{a}/(t+b)^{c}$, evaluated at the
25-year return period for a storm duration equal to the time of
concentration. The curve number (antecedent moisture condition II) was
assigned from the catchment hydrologic soil group, derived from
SoilGrids 250~m clay and sand fractions \cite{Poggio2021} via a
texture-to-group classification, combined with the biome-characteristic
land cover, using standard NRCS tables \cite{NRCS2004}.

For each catchment the design hyetograph was constructed by
differentiating the fitted cumulative Huff curve, converted to direct
runoff with the SCS-CN loss model (initial-abstraction ratio 0.2), and
convolved with the SCS dimensionless unit hydrograph (peak factor 484)
to obtain the outlet hydrograph. This was done twice. Once with the
original Huff (1967) Q1 reference curve and once with the
corresponding biome Q1 curve derived here. The relative change in
peak discharge, $\Delta Q_p$, and the shift in time-to-peak,
$\Delta t_p$, were  then recorded. Because the loss model and unit hydrograph
are identical in the two runs, systematic errors in curve number and
time of concentration largely cancel, isolating the effect of the
temporal pattern. The complete set of governing equations, the input
datasets and their spatial fields, the catchment-eligibility criteria,
and the curve-number assignment are documented in the Supplementary
Material.

\subsection{The Huff Curves BR Atlas: an interactive web application}
\label{sec:webatlas}

To maximise the practical impact and accessibility of the results, we
developed a free, browser-based interactive application — the
\emph{Huff Curves BR Atlas} — that exposes the complete station-level
database derived in this study without requiring any software
installation or programming expertise. The application is
available at \url{https://marcusnobrega-eng.github.io/Huff_Curves_BR},
and the source code and data are
archived together with the analysis pipeline (see the Open Research
Section). A full description of the application interface, analytical
outputs, and design-storm module is provided in the Supplementary
Material (Section~S8, Fig.~S3).

\section{Results}
\label{sec:results}

\subsection{Station processing and event characteristics}

A total of 290,164 qualifying events were extracted from the 1,045
accepted stations (see Section~\ref{sec:data} for quality-filter details), corresponding to a median of 265 events per
station (range: 8–777). Table~\ref{tab:events} summarises event
characteristics by dominant quartile. Q1 events account for 43.6\%
of all events; Q4 events are the least frequent at 13.1\%. Q1 events
are systematically shorter (median duration 9.0~h vs.\ 13.0~h for
Q4) and more intense (median peak intensity 20.4~mm~h$^{-1}$ vs.\
15.0–16.0~mm~h$^{-1}$ for Q2–Q4). Median event volumes are nearly
identical across quartiles (26.0–27.8~mm), confirming that quartile
assignment captures the \emph{timing} of rainfall rather than its
total depth.

\begin{table}[H]
  \centering
  \caption{Event characteristics by dominant quartile. Values are
    median (interquartile range in parentheses) unless noted.
    All events use IETD = 6~h and minimum storm total = 12.7~mm.
    The bottom panel shows the quartile share (\%) of the top-10\%
    most extreme events at each station, defined separately by
    storm volume and by peak 15-min intensity (29{,}575 events each;
    dash in the All column because shares sum to 100\%).}
  \label{tab:events}
  \begin{tabular}{lrrrrr}
    \toprule
    & Q1 & Q2 & Q3 & Q4 & All \\
    \midrule
    $N$ events      &
      126{,}565 & 70{,}396 & 55{,}132 & 38{,}071 & 290{,}164 \\
    \% of events    &
      43.6 & 24.3 & 19.0 & 13.1 & 100.0 \\
    \midrule
    Volume (mm)     &
      26.0 & 27.6 & 27.8 & 27.0 & 26.8 \\
                    &
      (18.0–40.8) & (18.8–44.0) & (18.8–45.0) & (18.0–49.6) & (18.2–43.2) \\
    Duration (h)    &
      9.0 & 10.5 & 12.0 & 13.0 & 10.0 \\
                    &
      (5.5–15.0) & (5.8–18.0) & (7.3–19.0) & (8.0–22.0) & (6.0–17.3) \\
    Avg intensity (mm~h$^{-1}$) &
      3.06 & 2.94 & 2.53 & 2.39 & 2.82 \\
    Max intensity (mm~h$^{-1}$) &
      20.4 & 16.0 & 15.0 & 16.0 & 17.6 \\
    \midrule
    \multicolumn{6}{l}{\textit{Seasonal Q1 fraction (\%)}} \\
    DJF (summer) & \multicolumn{4}{l}{45.4\% Q1 \quad (112{,}800 events total)} & \\
    MAM (autumn) & \multicolumn{4}{l}{44.1\% Q1 \quad (71{,}513 events total)} & \\
    JJA (winter) & \multicolumn{4}{l}{33.6\% Q1 \quad (31{,}480 events total)} & \\
    SON (spring) & \multicolumn{4}{l}{44.8\% Q1 \quad (74{,}371 events total)} & \\
    \midrule
    \multicolumn{6}{l}{\textit{Quartile share of extreme events (\%, top-10\% per station)}} \\
    By storm volume       & 39.4 & 25.8 & 21.3 & 13.5 & {—} \\
    By peak intensity     & 51.3 & 22.7 & 14.8 & 11.2 & {—} \\
    \bottomrule
  \end{tabular}
\end{table}

The seasonal distribution of events reveals a pronounced summer
maximum: DJF contributes 38.9\% of all events, compared with only
10.8\% in JJA (Table~\ref{tab:events}). The Q1 fraction varies
seasonally from 45.4\% (DJF) to 33.6\% (JJA), the latter reflecting
an increased role of organised frontal systems in austral winter.
March records the highest monthly Q1 fraction (46.6\%) and August
the lowest (32.7\%).

\begin{figure}[H]
  \centering
  \includegraphics[width=\linewidth]{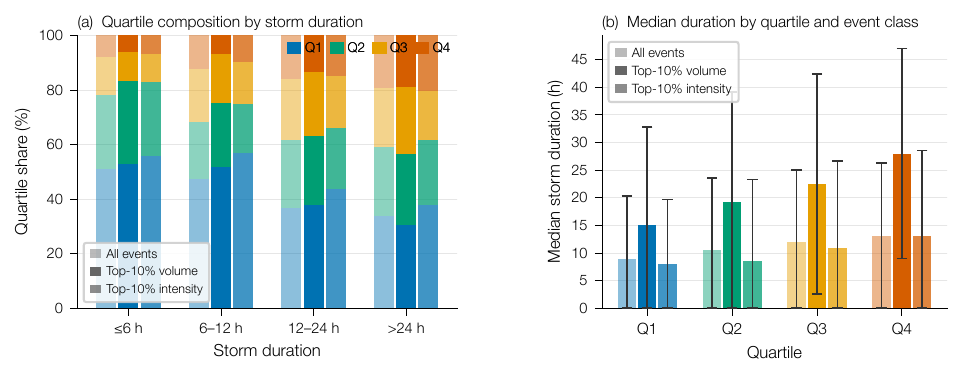}
  \caption{(a)~Quartile composition (\%) of storm events by duration
    for all events (light), the top-10\% most voluminous events per
    station (solid), and the top-10\% most intense events per station
    (medium shade). Within each duration class, three stacked bars are
    shown left to right for the three event classes.
    (b)~Median storm duration (h) by Huff quartile for the same three
    event classes; error bars show $\pm$1 standard deviation.
    Volume-extreme events (top-10\% by storm depth) are substantially
    longer than typical events of the same quartile (e.g., Q4 median:
    28~h vs.\ 13~h), explaining their more uniform quartile
    distribution; intensity-extreme events have similar or shorter
    durations, consistent with their heightened Q1 concentration.}
  \label{fig:extreme_duration}
\end{figure}

We also examined which quartile captures the most extreme events at
each station and how storm duration mediates that relationship
(Fig.~\ref{fig:extreme_duration}, Table~\ref{tab:events}). Storm
duration is itself strongly correlated with quartile: across all
events, Q1 accounts for 50.9\% of short storms ($\leq$6~h) but only
33.8\% of long storms ($>$24~h), with Q2–Q4 gaining share
monotonically as duration increases (Fig.~\ref{fig:extreme_duration}a).
This gradient reflects the convective origin of front-loaded events
and the stratiform or frontal character of later-peaking events.

When the analysis is restricted to the top 10\% of events by storm
volume within each station (29{,}575 events), Q1 retains the
plurality but its share decreases to 39.4\% (Q2: 25.8\%; Q3:
21.3\%; Q4: 13.5\%), compared with 43.6\%, 24.3\%, 19.0\%, and
13.1\% across all events. This shift is largely a duration effect:
volume extremes are substantially longer storms than typical events of
the same quartile (median durations of 15, 19, 22, and 28~h for
Q1–Q4, versus 9, 10.5, 12, and 13~h across all events;
Fig.~\ref{fig:extreme_duration}b), and, within each duration
bracket, the quartile composition of volume extremes closely mirrors
that of all events. The apparent shift towards Q2 and Q3 among
high-volume extremes is therefore not a change in storm character but
a consequence of large-volume events requiring longer storm durations,
which naturally reduces Q1 dominance. The pattern is reversed for
peak rainfall intensity: among the top-10\% most intense events at
each station, Q1 accounts for 51.3\% (Q2: 22.7\%; Q3: 14.8\%; Q4:
11.2\%), because the most intense storms are concentrated in the
$\leq$6~h bracket where Q1 is most dominant. These contrasting
patterns carry a direct implication for design: the choice of Huff
quartile matters most for infrastructure sensitive to peak discharge
(e.g., urban storm drains and culverts), while for volume-controlled
applications such as reservoir flood routing, all four quartile curves
remain relevant across the full range of storm durations.

\subsection{Diurnal cycle of event initiation}
\label{sec:diurnal}

\begin{figure}[H]
  \centering
  \includegraphics[width=\columnwidth]{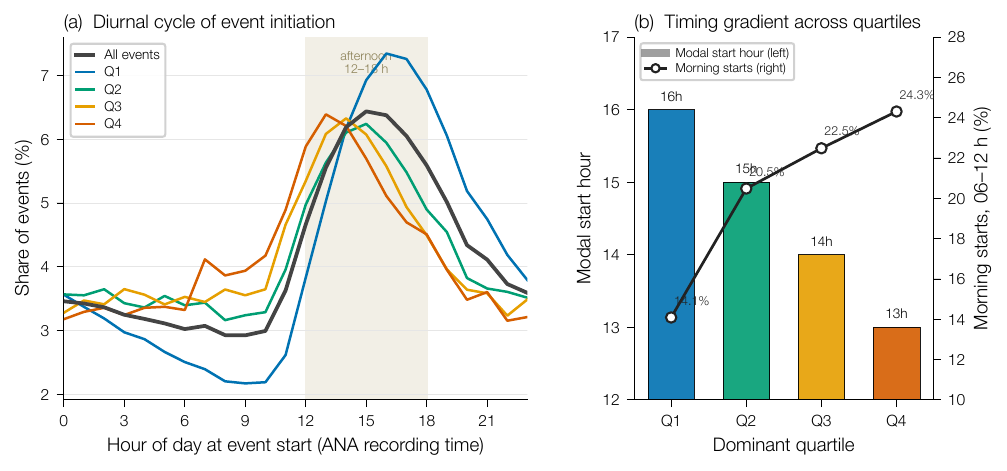}
  \caption{Diurnal cycle of rainfall-event initiation
    (290{,}164 events). (a) Hour-of-day distribution of event start
    times for all events (grey) and for each dominant quartile
    (Q1–Q4, coloured); the shaded band marks the afternoon window
    (12–18~h). (b) Timing gradient across quartiles: bars give the
    modal start hour (left axis), and the line gives the fraction of
    events beginning in the morning, 06–12~h (right axis). Hours follow
    the ANA telemetric recording convention; the monotonic ordering of
    the quartiles is invariant to any uniform time-zone offset. First-
    quartile (front-loaded) storms peak latest in the day and are least
    likely to begin in the morning, consistent with afternoon
    convective forcing.}
  \label{fig:diurnal}
\end{figure}

Because the ANA records preserve the timestamp of every event, we also
examined the hour of day at which events begin
(Fig.~\ref{fig:diurnal}). Event initiation follows a pronounced diurnal
cycle: across all 290,164 events, the rate of initiation is lowest in
the late morning and rises to a sharp maximum at a modal start hour of
15~h, so that 35.3\% of events begin in the afternoon window
(12–18~h) against only 18.6\% in the morning (06–12~h)
(Fig.~\ref{fig:diurnal}a). The absolute clock hours follow the
recording convention of the ANA telemetric network and are not
adjusted for the several time zones spanned by the country; the
\emph{relative} timing across quartiles described below is, however,
invariant to any uniform offset.

The diurnal signal varies systematically and monotonically with the
dominant quartile (Fig.~\ref{fig:diurnal}b). The modal start hour
advances by a full hour per quartile, from 16~h for Q1 to 15~h, 14~h,
and 13~h for Q2, Q3, and Q4 respectively, while the fraction of events
beginning in the morning rises in the same order, from 14.1\% for Q1 to
20.5\%, 22.5\%, and 24.3\% for Q2–Q4. First-quartile (front-loaded) storms therefore have the latest
diurnal initiation time and are the least likely to begin in the
morning; fourth-quartile (back-loaded) storms initiate earliest in
the day and are the most morning-prone. This ordered gradient links
the quartile classes to a common diurnal driver and is interpreted
further in Section~\ref{sec:mechanism}.

\subsection{National Huff curves}
\label{sec:national}

\begin{figure}[H]
  \centering
  \includegraphics[width=\columnwidth]{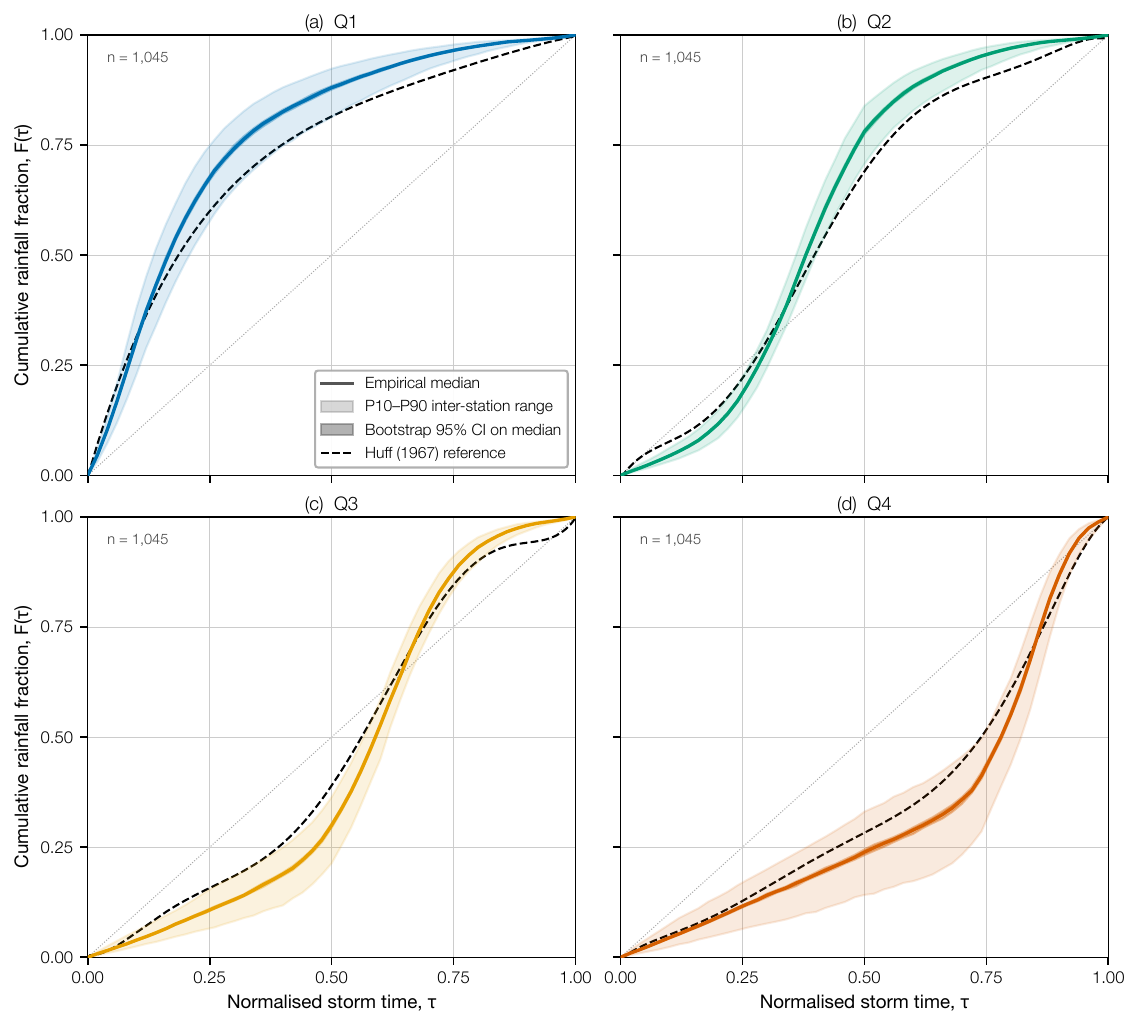}
  \caption{National empirical Huff curves for Brazil derived from
    1{,}045 rain-gauge stations (290{,}164 events, 2010–2025).
    Panels (a)–(d): quartiles Q1–Q4. Solid coloured line: national
    median empirical curve. Light shaded band: P10–P90 inter-station
    range. Darker shaded band: bootstrap 95\% confidence interval on
    the national median ($B=2{,}000$ resamples). Dashed black line:
    Huff (1967) reference curve for that quartile (Illinois, USA).
    Dotted diagonal: uniform temporal distribution ($F = \tau$).}
  \label{fig:national}
\end{figure}

Fig.~\ref{fig:national} shows the national empirical Huff curves for
Q1–Q4, including the inter-station P10–P90 range, the bootstrap 95\%
confidence interval on the national median, and the Huff (1967)
reference. Q1 is the dominant quartile at 986 of the 1,045 accepted
stations (94.4\%); Q2, Q3, and Q4 account for 2.0\%, 0.9\%, and
2.8\%, respectively.

The national Q1 median curve lies above the Huff (1967) reference
throughout $\tau \in [0.1, 0.9]$, indicating that Brazilian Q1
storms accumulate a larger fraction of their total depth earlier in
the storm than the Illinois reference. The national Q1 MAE is 0.045
and $D_\mathrm{max}$ is 0.097; at the median $\tau = 0.5$, the
empirical curve reaches $F = 0.86$ compared with $F = 0.82$ for the
reference, a difference of four percentage points of storm total.
Bootstrap uncertainty is small: the mean 95\% CI width across
$\tau$ points is 0.006 for Q1, confirming that the national median
is precisely estimated despite inter-station variability.

For Q2, Q3, and Q4, the national median curves lie close to their
respective Huff (1967) references (national MAE: 0.042, 0.042, and
0.036 respectively), with the largest absolute deviations occurring
in the middle of the storm ($\tau \approx 0.4$–$0.6$). The broader
P10–P90 ranges for Q2–Q4 reflect higher intra-station variability in
storms where the peak is not concentrated in the first quarter.

\subsection{Spatial variability of Huff curves across Brazilian biomes}
\label{sec:biome}

\begin{figure}[H]
  \centering
  \includegraphics[width=\columnwidth]{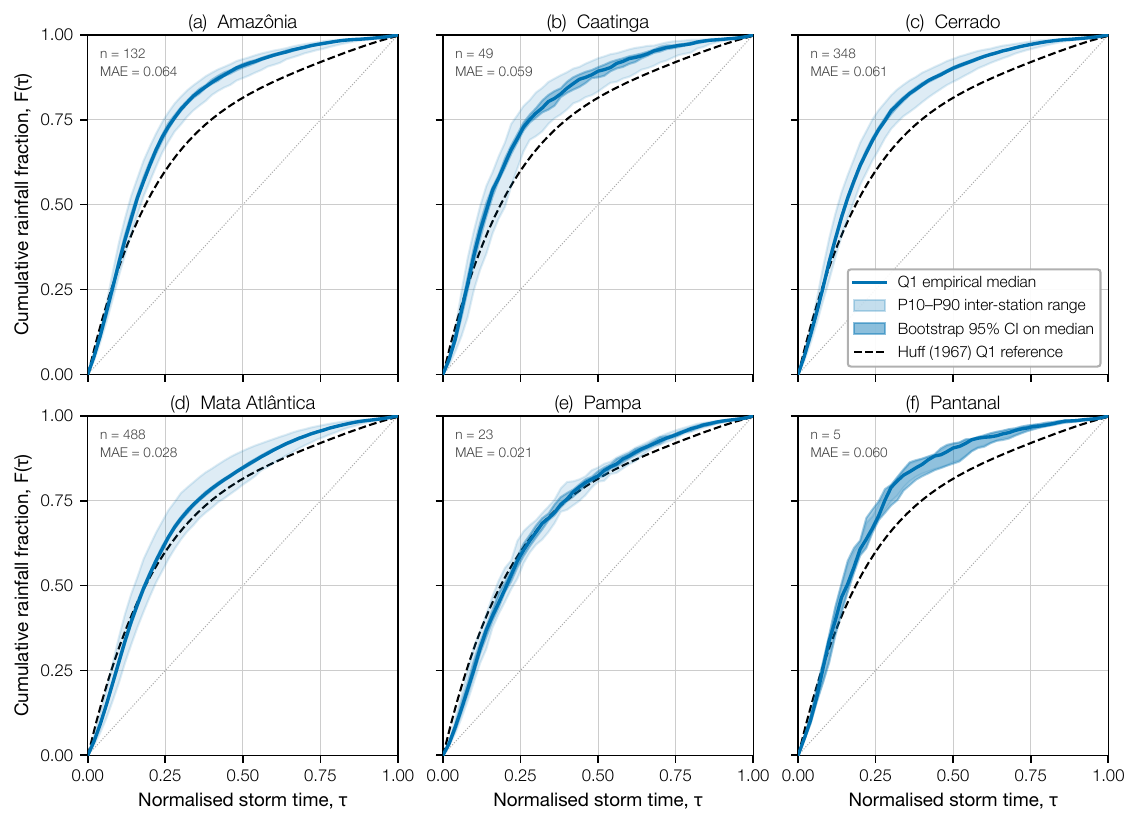}
  \caption{Q1 empirical Huff curves by IBGE biome. Layout as in
    Fig.~\ref{fig:national}. Station count ($n$) and MAE relative
    to the Huff (1967) Q1 reference are annotated in each panel.
    The Pantanal (f) result is based on $n=5$ stations and should
    be interpreted cautiously.}
  \label{fig:biome}
\end{figure}

Fig.~\ref{fig:biome} and Table~\ref{tab:biome} present Q1 curve
characteristics at the biome scale. Q1 is the dominant quartile in
all six biomes, with station fractions ranging from 88.7\% in the
Mata Atlântica to 100\% in the Pantanal. The agreement with the Huff
(1967) reference, however, varies systematically across the biomes
in a manner that closely tracks their underlying storm climatology.

\begin{table}[H]
  \centering
  \caption{Biome-level summary of Q1 Huff curve characteristics.
    $n_\mathrm{sta}$: number of accepted stations.
    $N_\mathrm{ev}$: total qualifying events (IETD~=~6~h;
    minimum depth~=~12.7~mm).
    $\tilde{r}$: median station record length (yr).
    $f_\mathrm{Q1}$: fraction of stations for which Q1 is
    the dominant (modal) quartile, i.e.\ the most frequent
    event-timing class at that station.
    MAE: mean absolute error of the biome Q1 median curve
    $\hat{F}(\tau)$ relative to the Huff (1967) Q1 reference
    $F_\mathrm{ref}(\tau)$, averaged over 20 equally spaced values
    of normalised storm time $\tau \in (0, 1]$:
    $\mathrm{MAE} = \tfrac{1}{20}\sum_{j}|\hat{F}(\tau_j)
    - F_\mathrm{ref}(\tau_j)|$.
    $D_\mathrm{max}$: maximum absolute deviation
    $\max_\tau|\hat{F}(\tau) - F_\mathrm{ref}(\tau)|$.
    95\%~CI~width: bootstrap 95\% confidence interval width of
    $\hat{F}(\tau)$ at $\tau = 0.5$ (B~=~2{,}000 station resamples),
    given in parentheses.
    Pantanal ($n_\mathrm{sta}=5$): interpret with caution.}
  \label{tab:biome}
  \setlength{\tabcolsep}{4pt}
  \begin{tabular}{lrrrrrrr}
    \toprule
    Biome & $n_\mathrm{sta}$ & $N_\mathrm{ev}$ & $\tilde{r}$ (yr) &
      $f_\mathrm{Q1}$ (\%) & MAE & $D_\mathrm{max}$ &
      95\% CI width \\
    \midrule
    Amazônia        & 132 & 50{,}299 & 10.5 & 99.2 & 0.063 & 0.121 & (0.011) \\
    Caatinga        &  49 &  6{,}491 &  7.3 & 98.0 & 0.059 & 0.112 & (0.029) \\
    Cerrado         & 348 & 85{,}521 & 10.4 & 99.7 & 0.060 & 0.114 & (0.007) \\
    Mata Atlântica  & 488 &141{,}356 & 10.8 & 88.7 & 0.028 & 0.042 & (0.007) \\
    Pampa           &  23 &  4{,}913 &  7.4 & 95.7 & 0.019 & 0.062 & (0.019) \\
    Pantanal        &   5 &  1{,}584 & 16.0 & 100.0 & 0.058 & 0.115 & (0.059) \\
    \midrule
    \textbf{National} & \textbf{1{,}045} & \textbf{290{,}164} &
      \textbf{10.6} & \textbf{94.4} & \textbf{0.045} &
      \textbf{0.097} & \textbf{(0.006)} \\
    \bottomrule
  \end{tabular}
\end{table}

The Amazônia biome (132 stations, 50,299 events) is strongly
Q1-dominant at 99.2\% of stations, but its median curve shows the
largest positive departure from the Illinois reference of any biome
(MAE = 0.063, $D_\mathrm{max}$ = 0.121), lying well above it
throughout the storm. This reflects the high convective available
potential energy of the equatorial Amazon, where deep convective
cells release the bulk of their rainfall rapidly after initiation.
The Caatinga (49 stations, 6,491 events) behaves similarly, with a
Q1 fraction of 98.0\% and comparable departures from the reference
(MAE = 0.059, $D_\mathrm{max}$ = 0.112); despite the small station
count, its Q1 dominance is robust across all tested inter-event time
definitions ($\geq$~95.9\% for IETD~$\geq$~4~h), consistent with the
rare but intense and short-lived storms produced by the seasonal
migration of the Intertropical Convergence Zone. The Cerrado,
the largest biome in the dataset by event count (348 stations,
85,521 events), is the most uniformly Q1-dominant of the continental
biomes (99.7\%; MAE = 0.060, $D_\mathrm{max}$ = 0.114), reflecting
the near-exclusive role of afternoon thermally driven convection over
its flat interior plateau in producing consistently front-loaded mass
curves.

The two biomes most influenced by extratropical and orographic
forcing depart from this pattern in opposite directions. The Mata
Atlântica (488 stations, 141,356 events) has the lowest Q1 fraction
of all biomes at 88.7\%, yet paradoxically the best agreement with
the Huff reference (MAE = 0.028, $D_\mathrm{max}$ = 0.042; bootstrap
95\% CI width at $\tau = 0.5$ of 0.007). Both features stem from the
same cause: the diversity of storm types in this biome — orographic
rainfall along the Serra do Mar, sea-breeze-driven nocturnal events,
and cold fronts arriving from the south — generates a substantial
minority of Q2–Q4 storms and, for the Q1 population that remains, a
median curve closer to the more gradual Illinois shape. The Pampa
(23 stations, 4,913 events) shows the closest match to the Illinois
reference of any biome (MAE = 0.019, $D_\mathrm{max}$ = 0.062) while
retaining a high Q1 fraction of 95.7\%; its extratropical regime most
closely resembles the mid-latitude climate sampled by \citeA{Huff1967}.
Finally, the Pantanal (5 stations, 1,584 events) is Q1-dominant at
every station and every IETD value (MAE = 0.058, $D_\mathrm{max}$ =
0.115), although its small station count means these results are
indicative only.

\subsection{Spatial patterns of Huff curves}
\label{sec:maps}

\begin{figure}[H]
  \centering
  \includegraphics[width=\columnwidth]{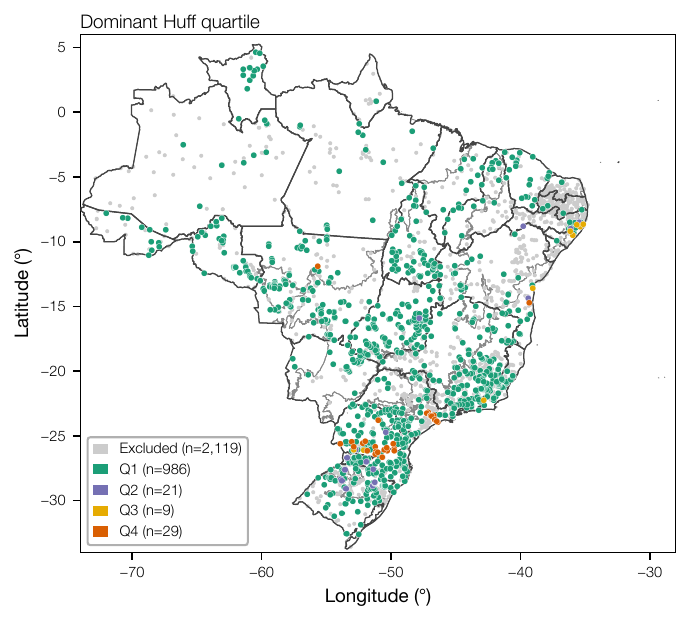}
  \caption{Spatial distribution of the dominant Huff quartile at
    each of the 1{,}045 accepted Brazilian rain-gauge stations.
    Grey circles ($n=2{,}119$) indicate stations excluded by
    quality control. Light and dark grey lines: IBGE biome and
    state boundaries, respectively.}
  \label{fig:map_q}
\end{figure}

The dominant quartile at each of the 1,045 accepted stations is mapped
in Fig.~\ref{fig:map_q}. Q1-dominant stations are distributed
essentially uniformly across the national territory, spanning the
equatorial Amazon, the semi-arid Northeast, the central Cerrado, and
the subtropical south without any pronounced regional gradient. This
spatial homogeneity is itself a key result: it indicates that the
front-loaded storm structure is not a feature of one particular
climate zone but a pervasive characteristic of Brazilian heavy
rainfall. The few stations classified as Q2, Q3, or Q4 do not form
large contiguous clusters; the only discernible spatial signal is a
modest concentration of Q4 (late-peaking) stations along the southern
Pampa and the coastal Mata Atlântica, precisely the two regions where
extratropical cold fronts most frequently produce longer, more
symmetric or back-loaded rainfall sequences.

\begin{figure}[H]
  \centering
  \includegraphics[width=\columnwidth]{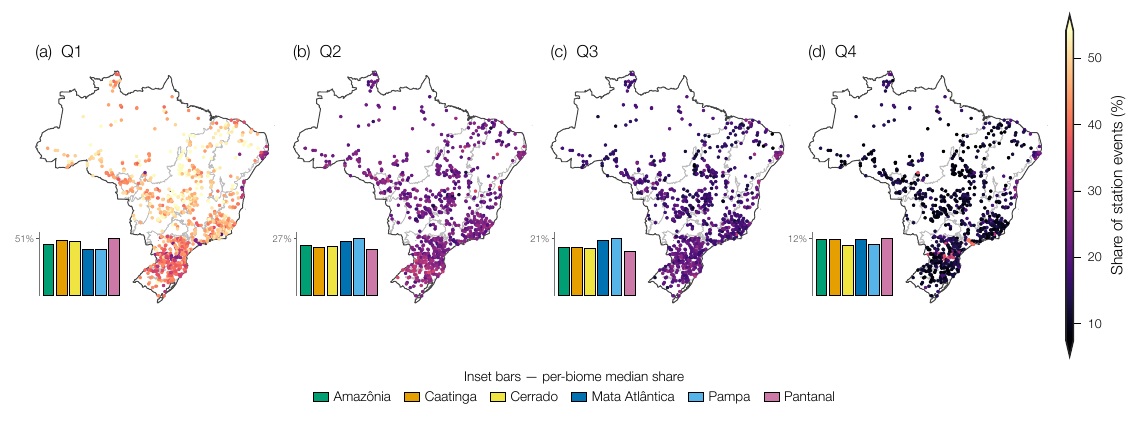}
  \caption{Per-station quartile composition of the event population.
    Maps (a)–(d) show, for each accepted station, the share of its
    events falling in the first, second, third, and fourth quartile,
    respectively, over biome (light grey) and national (dark grey)
    boundaries. A single colour scale is shared across the four panels
    (arrowed ends mark values beyond the 2nd–98th percentile of the
    pooled data, which are clipped), so the panels directly convey the
    magnitude ordering Q1~$>$~Q2~$>$~Q3~$>$~Q4. The bottom-left inset
    of each map gives the per-biome median share for that quartile,
    coloured and ordered as in the legend, with the panel maximum
    labelled.}
  \label{fig:quartile_percent}
\end{figure}

The dominant-quartile classification reduces each station to its modal
class and therefore conceals the underlying composition of the event
population. Mapping the full distribution, the share of each station's
events in every quartile (Fig.~\ref{fig:quartile_percent}), shows that
the composition is remarkably stable nationally but varies
systematically with biome in a way the dominant-quartile map cannot
resolve. The first-quartile share is highest in the convective biomes
(here defined as those dominated by thermally driven deep convection:
Pantanal, Caatinga, Cerrado, and Amazônia; \citeNP{Marengo2012,
	Zipser2006}), with biome-median values of 50.8\% (Pantanal), 49.6\%
(Caatinga), 48.3\% (Cerrado), and 46.1\% (Amazônia), and lowest in
the two extratropical biomes (Mata Atlântica and Pampa, where frontal
and orographic forcing is substantial), 41.2\% and 41.4%
respectively. The deficit in Q1 in the south is taken up almost entirely by the
second and third quartiles, whose combined biome-median share rises
from about 40\% in the convective interior to 45–47\% in the Mata
Atlântica and Pampa, while the fourth-quartile share remains nearly
constant (10.6–11.9\%) across all biomes. In other words, the
north-to-south transition from convective to frontal regimes is
expressed not as a switch in the dominant quartile, which remains Q1
almost everywhere, but as a gradual transfer of probability mass from
the first quartile into the mid-storm quartiles.

\begin{figure}[H]
  \centering
  \includegraphics[width=\columnwidth]{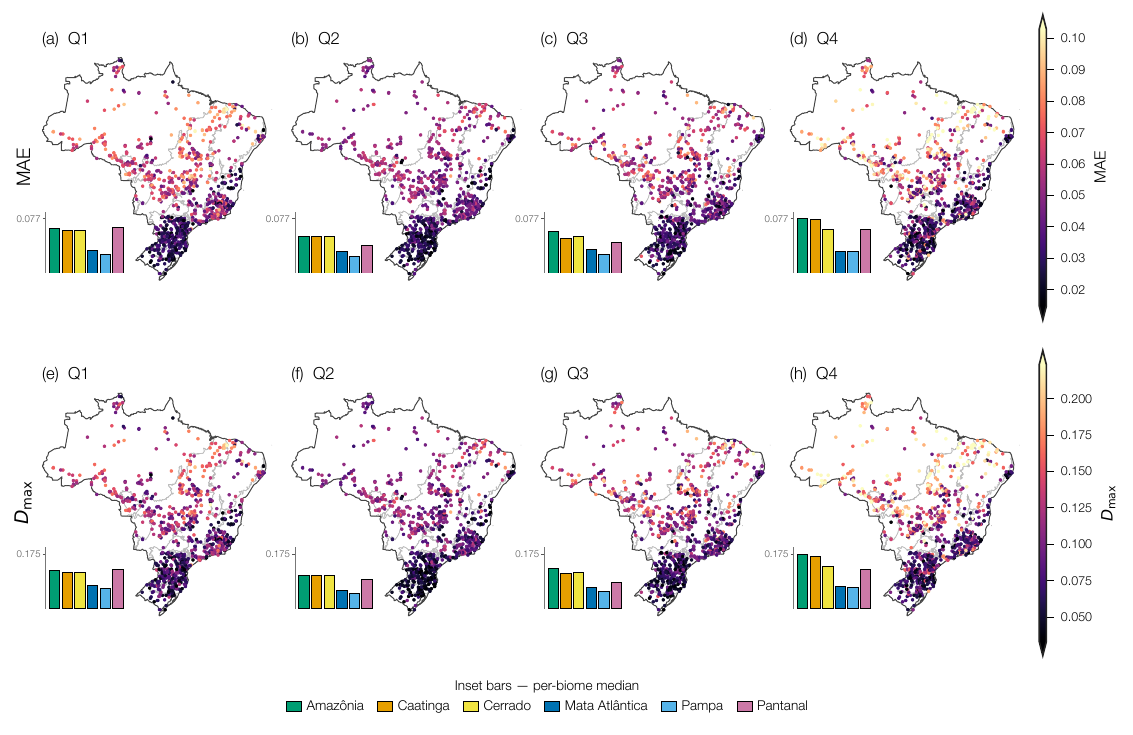}
  \caption{Per-quartile spatial distribution of goodness-of-fit metrics
    for the empirical Huff curves relative to the Huff (1967) Illinois
    reference. Top row (a–d): mean absolute error (MAE) for quartiles
    Q1–Q4; bottom row (e–h): maximum absolute deviation
    ($D_\mathrm{max}$, the Kolmogorov–Smirnov statistic) for Q1–Q4.
    Each station is coloured by its own per-quartile metric, over biome
    (light grey) and national (dark grey) boundaries. Within each row
    the colour scale is shared across all four quartiles and the
    arrowed (extended) ends indicate values beyond the 2nd–98th
    percentile of the pooled data, which are clipped; brighter (yellow)
    tones denote larger departures from the reference. The inset bar
    chart in the lower-left of each map gives the median metric for
    each of the six IBGE biomes, ordered and coloured as in the legend
    (Amazônia, Caatinga, Cerrado, Mata Atlântica, Pampa, Pantanal);
    the inset uses a common vertical scale within each row, whose
    maximum is labelled.}
  \label{fig:map_mae}
\end{figure}

The spatial distribution of the goodness-of-fit metrics, computed
separately for each quartile (Fig.~\ref{fig:map_mae}), reveals two
superimposed and consistent patterns. The first is a coherent
north–south geographic gradient that is present in every quartile: the
lowest MAE and $D_\mathrm{max}$ values, the closest agreement with the
Huff (1967) reference, are concentrated in the south and southeast,
consistent with the climatically more Illinois-like conditions of the
Pampa and the southern Mata Atlântica, whereas the highest values occur
in the Amazon interior and parts of the Caatinga, where intense
tropical convection drives the empirical curves furthest from the
reference. At the biome level the median MAE for Mata Atlântica and
Pampa remains below 0.035 for all four quartiles, against 0.05–0.08 for
Amazônia, Caatinga, and Pantanal.

The second pattern is a systematic dependence on the quartile itself.
The empirical curves match the reference best for the second quartile
and worst for the fourth: nationally, the median MAE rises from 0.043
(Q2) to 0.047 (Q4) and the median $D_\mathrm{max}$ from 0.085 (Q2) to
0.103 (Q4), with Q1 and Q3 intermediate. This Q4 deterioration is
strongest precisely in the convective biomes: in Amazônia the median
Q4 $D_\mathrm{max}$ reaches 0.175, by far the largest departure of any
biome–quartile combination, compared with 0.108 for its Q2. The
back-loaded fourth-quartile pattern is therefore both the rarest
(13.1\% of events; Section~\ref{sec:national}) and the least
well-described by the Illinois reference in the regions where
convection dominates. The two effects compound: agreement with the
Huff (1967) curves is highest for the early- and mid-peaking storms of
the extratropical south, and lowest for the late-peaking storms of the
deep tropics.

\subsection{IETD sensitivity}
\label{sec:ietd}

\begin{figure}[H]
  \centering
  \includegraphics[width=\columnwidth]{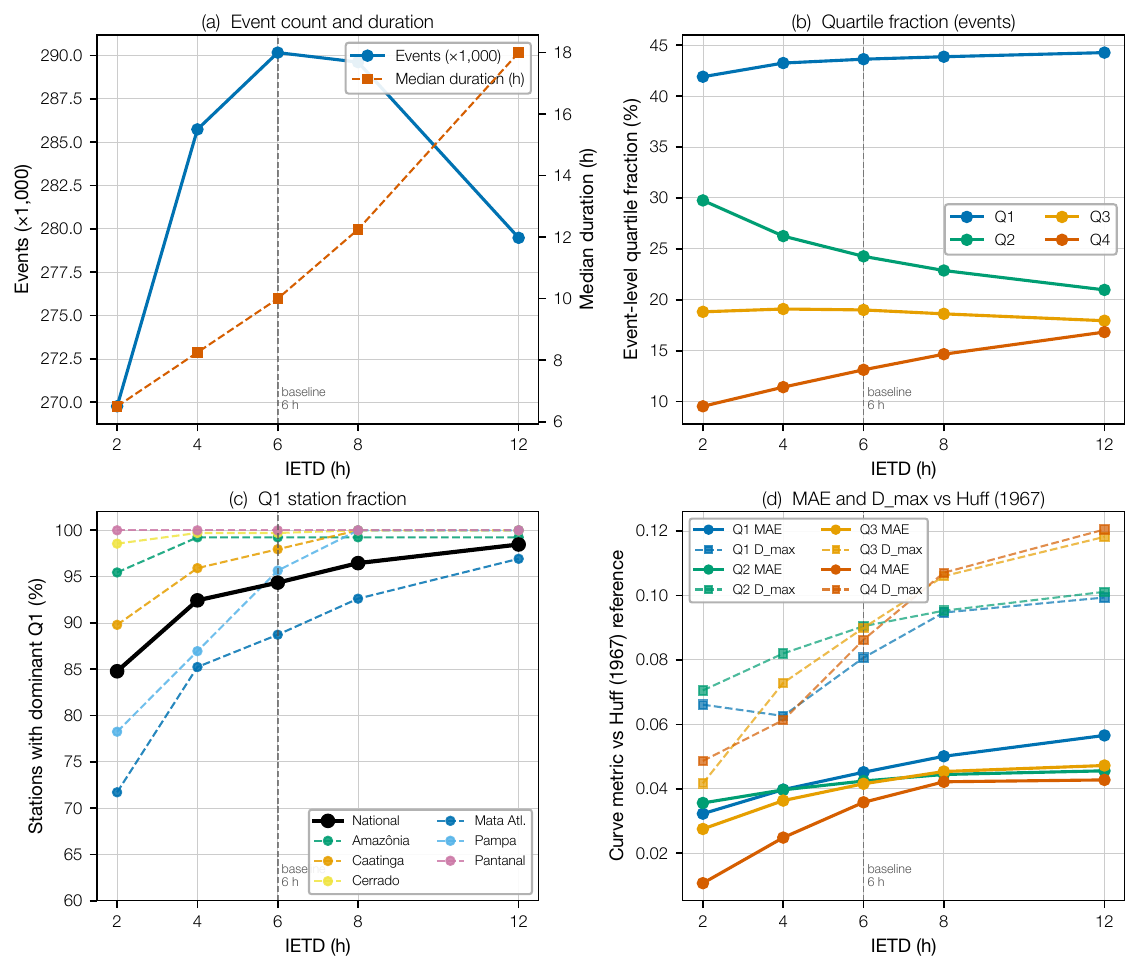}
  \caption{IETD sensitivity analysis across $\{2, 4, 6, 8, 12\}$~h.
    (a) Total qualifying event count (solid, left axis) and median
    event duration (dashed, right axis). (b) Event-level quartile
    fraction (\%). (c) Fraction of stations with Q1 as dominant
    quartile, shown nationally (thick black line) and by biome.
    (d) MAE (solid) and $D_\mathrm{max}$ (dashed) of empirical Q1–Q4
    national median curves relative to the Huff (1967) reference.
    Vertical dashed grey line marks the baseline IETD = 6~h used
    in the main analysis.}
  \label{fig:ietd}
\end{figure}

Table~\ref{tab:ietd} and Fig.~\ref{fig:ietd} summarise the
sensitivity analysis across five IETD values, which together
demonstrate that the principal conclusions of this study are robust
to the choice of inter-event separation.

\begin{table}[H]
  \centering
  \caption{IETD sensitivity analysis. Baseline (IETD = 6~h) in bold.
    ``Agreement'' = fraction of the 1{,}045 baseline-OK stations
    retaining the same dominant quartile as the 6-h baseline.
    MAE and $D_\mathrm{max}$ computed for the national Q1 median
    empirical curve relative to the Huff (1967) reference.}
  \label{tab:ietd}
  \begin{tabular}{rrrrrr}
    \toprule
    IETD (h) & $N$ events & Med dur (h) & Q1 sta (\%) &
      Agreement & Q1 MAE \\
    \midrule
    2  & 269{,}768 &  6.5 & 84.8 & 88.1\% & 0.032 \\
    4  & 285{,}737 &  8.3 & 92.4 & 96.3\% & 0.040 \\
    \textbf{6} & \textbf{290{,}164} & \textbf{10.0} & \textbf{94.4} &
      \textbf{—} & \textbf{0.045} \\
    8  & 289{,}615 & 12.3 & 96.5 & 97.1\% & 0.050 \\
    12 & 279{,}484 & 18.0 & 98.5 & 95.7\% & 0.057 \\
    \bottomrule
  \end{tabular}
\end{table}

The station-level dominant-quartile classification is highly stable.
Across all five IETD values from 2 to 12~h, 84.7\% of the 1,045
stations retain the same dominant quartile, and agreement with the
6-h baseline reaches 96.3\% at IETD = 4~h and 97.1\% at IETD = 8~h,
so that the $\pm$2~h bracket around the chosen value changes the
classification of only about 3\% of stations. The sensitivity that
does exist is concentrated in the two biomes with the most diverse
storm populations: the Mata Atlântica and the Pampa, whose Q1
fractions rise from 71.7\% and 78.3\% at IETD = 2~h to 96.9\% and
100\% at 12~h. The Amazônia, Cerrado, and Caatinga biomes, dominated
by short convective cells, are essentially invariant above IETD = 4~h.

The total number of qualifying events is itself maximised at the
chosen value. Event count peaks at 290,164 for IETD = 6~h and declines
on either side: to 269,768 at 2~h, as short inter-event gaps split
storms into sub-events that individually fall below the 12.7~mm
minimum depth, and to 279,484 at 12~h, as excessive merging reduces
the number of discrete storms. This interior maximum provides an
independent, data-driven justification for the 6-h choice beyond the
literature precedent discussed in Section~\ref{sec:event_extraction}.

Finally, the agreement between the empirical curves and the Huff
(1967) reference degrades monotonically as the IETD lengthens: the
national Q1 MAE increases from 0.032 at IETD = 2~h to 0.057 at 12~h.
The mechanism is physical: at longer separations, merged events
increasingly incorporate stratiform trailing rainfall that flattens
the early rise of the cumulative mass curve and so moves it away from
the sharply front-loaded Huff Q1 shape. This behaviour provides an
objective, rather than merely conventional, basis for preferring
short-to-moderate IETD values in this application, with IETD = 6~h
representing the best available compromise between maximising the
qualifying event count and preserving fidelity to the reference curve.

\subsection{Impact on design peak discharge}
\label{sec:hydro_results}

\begin{figure}[H]
  \centering
  \includegraphics[width=0.85\columnwidth]{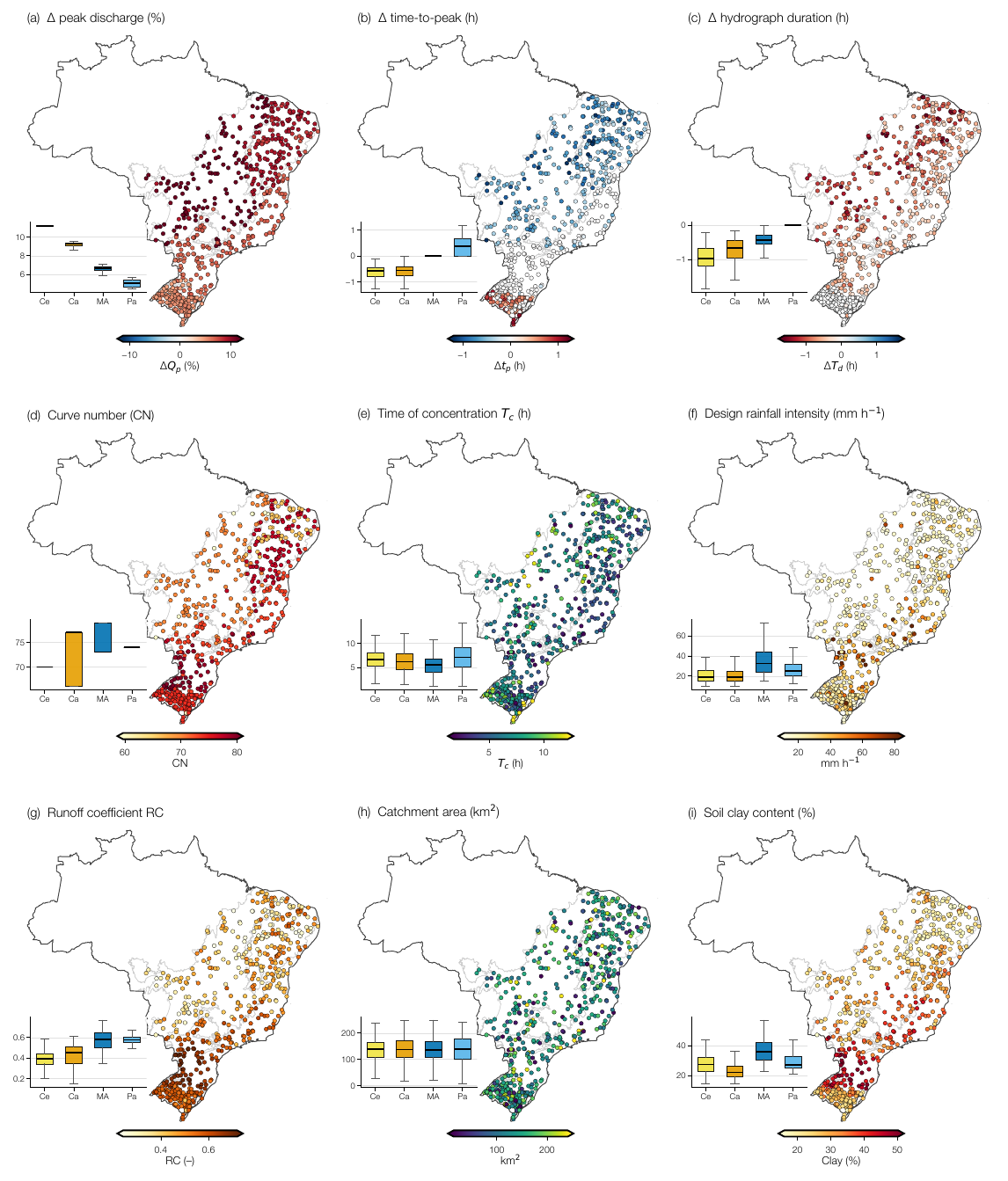}
  \caption{Hydrological impact of replacing the original Huff (1967) Q1
    curve with the locally derived biome Q1 curve, from a controlled
    design-hydrograph experiment on 579 small, SCS-eligible HydroBASINS
    level-12 catchments.
    \textit{Row 1 — response variables:}
    (a)~Change in design peak discharge $\Delta Q_p$ (\%);
    (b)~time-to-peak shift $\Delta t_p$ (h; negative = earlier peak);
    (c)~change in total hydrograph duration $\Delta T_d$ (h), defined
    as the time for discharge to return to 1\% of its peak value.
    \textit{Row 2 — rainfall and routing descriptors:}
    (d)~SCS curve number CN (antecedent moisture condition II);
    (e)~time of concentration $T_c$ (h) estimated from the
    Kirpich (1940) formula;
    (f)~25-year design rainfall intensity (mm~h$^{-1}$) from the
    gridded Sherman IDF surfaces of Gomes Junior (2025).
    \textit{Row 3 — catchment physical properties:}
    (g)~runoff coefficient $RC = Q_\mathrm{runoff}/P$, identical for
    both Huff patterns by the SCS-CN model (see Section~\ref{sec:limitations});
    (h)~catchment drainage area (km$^2$);
    (i)~soil clay fraction (\%) from SoilGrids 250~m
    (Poggio et al., 2021).
    In each panel, catchments are plotted as coloured dots; the
    biome-level distribution is summarised by the boxplot inset
    (Ce~=~Cerrado, Ca~=~Caatinga, MA~=~Mata~Atlântica, Pa~=~Pampa;
    boxes span the interquartile range; whiskers the 5th–95th
    percentiles).
    Catchment geometry, CN, $T_c$, design depth, and storm duration are
    held identical between the two hydrograph runs, isolating the effect
    of the temporal rainfall pattern.}
  \label{fig:hydro}
\end{figure}

The design-hydrograph experiment quantifies the engineering consequence
of replacing the imported Illinois reference with the locally derived
curves (Fig.~\ref{fig:hydro}). Across the 579 SCS-eligible catchments
for which all inputs were available, adopting the updated biome Q1 curve
\emph{increases} the design peak discharge relative to the original Huff
(1967) Q1 curve at every catchment, by a national median of 7.7\%
(interquartile range 6.2–9.5\%). The sign is consistent and physically
expected: because the Brazilian Q1 curves are systematically steeper
than the Illinois reference (Section~\ref{sec:national}), they deliver a
larger fraction of the storm depth in the first quarter, producing a
flashier hydrograph with a higher and earlier peak.

The magnitude of the effect follows the same convective-to-frontal
gradient seen throughout the analysis (Fig.~\ref{fig:hydro}b). The
increase is largest in the convective interior — a median of 11.2\% in
the Cerrado and 9.3\% in the Caatinga — and smallest in the more
Illinois-like extratropical south, 6.7\% in the Mata Atlântica and 5.1\%
in the Pampa. In other words, the imported curves underestimate design
peaks most severely precisely where the local storm climatology departs
furthest from Illinois, exactly as the goodness-of-fit maps
(Fig.~\ref{fig:map_mae}) would predict. The time-to-peak shifts
correspondingly (Fig.~\ref{fig:hydro}c): the more front-loaded updated
curves advance the peak by a median of about 0.6~h in the Cerrado and
Caatinga, leave it essentially unchanged in the Mata Atlântica, and
delay it slightly (about 0.4~h) in the Pampa, whose Q1 curve is the
closest of all biomes to the reference. The tight within-biome spread of
$\Delta Q_p$, obtained despite real catchment-to-catchment variation in
soil-derived curve number, time of concentration, and design depth,
indicates that the result is governed by the curve shape rather than by
the other model inputs.

\section{Discussion}
\label{sec:discussion}

\subsection{Mechanism of Q1 dominance}
\label{sec:mechanism}

Two distinct measures of Q1 prevalence must be carefully distinguished
before comparing our results with the literature, because they answer
different questions and take very different numerical values. The
\emph{event-level} Q1 fraction is the proportion of individual storms
classified as first-quartile, and equals 43.6\% nationally
(Table~\ref{tab:events}). The \emph{station-level} Q1 dominance is the
proportion of stations whose single most frequent quartile is Q1, and
equals 94.4\% (Section~\ref{sec:national}). The two differ because, at
the great majority of stations, Q1 is the modal class even though it
accounts for well under half of that station's events: with a typical
split on the order of 44\% Q1 against 24\%, 19\%, and 13\% for Q2–Q4,
Q1 is almost always the largest single group without being an absolute
majority. The 94.4\% figure therefore does \emph{not} imply that 94\%
of Brazilian storms are front-loaded; it states that front-loaded
storms are the most common type at 94\% of gauges.

Interpreted on a like-for-like basis, our results are consistent with
the international literature at the event level while extending it
substantially in spatial scope. Our event-level Q1 fraction of 43.6\%
is modestly higher than the 30\% reported by \citeA{Huff1967} for
Illinois and closely matches the 44.7–46.7\% found by
\citeA{SantaCatarina2021} in southern Brazil (see
Section~\ref{sec:comparison} for a full comparison). What is genuinely
new is the near-universal \emph{station-level} dominance of Q1 across a
continental and climatically diverse territory, a pattern that has not
previously been documented at national scale and that we propose
reflects a fundamental commonality in storm-generating mechanism.

Across most Brazilian biomes, rainfall is driven by deep moist
convection initiated by diurnal surface heating \cite{Marengo2012},
and the diurnal cycle of event initiation documented in
Section~\ref{sec:diurnal} (Fig.~\ref{fig:diurnal}) provides direct
empirical support for this mechanism. The afternoon maximum in event
initiation is the signature expected of thermally driven convection, in
which surface heating builds convective available potential energy
through the morning and triggers intense, relatively short storms
(median Q1 duration $\leq$~9~h) in the afternoon, concentrating the bulk
of storm depth in the first quarter of normalised time. Most
tellingly, the diurnal signal strengthens monotonically with the degree
of front-loading: first-quartile storms peak latest in the day and are
the least likely to begin in the morning, whereas back-loaded
fourth-quartile storms peak earliest and most often begin in the
morning. Because this ordering across quartiles is invariant to the
absolute clock convention of the ANA records, it constitutes robust
internal evidence that the front-loaded Q1 storms are the most direct
expression of afternoon convective forcing. The interpretation is
further consistent with the fact that South America hosts some of the
world's most intense continental thunderstorms \cite{Zipser2006},
matching the steep early rise of the Brazilian Q1 curves relative to
the Illinois reference.

The seasonal signal supports this interpretation: the Q1 fraction
drops from 45.4\% (DJF, austral summer, peak convective season)
to 33.6\% (JJA, austral winter). In winter, extratropical cold fronts
penetrate into southern Brazil and weaken the SACZ, shifting the storm
population towards more organised, longer-duration systems that
distribute rainfall more evenly in time, producing Q2 and Q3
patterns. The monthly trough in Q1 fraction (32.7\% in August)
coincides with maximum cold-front frequency in southern Brazil.

Biome differences are consistent with this framework. The Cerrado
(99.7\% Q1) experiences almost exclusively afternoon convective
storms over a flat interior plateau with high CAPE; the Mata Atlântica
(88.7\% Q1) has the most diverse storm population, including
orographic events along the escarpment, sea-breeze-driven nocturnal
rainfall, and frontal systems, all of which can produce non-Q1 mass
curves. The Pampa (95.7\% Q1), despite being the most extratropical
biome, still shows Q1 dominance because the long-duration frontal
events that might produce Q2–Q4 patterns are filtered out by the
12.7~mm minimum depth and the 6-h IETD: only the most intense cells
within frontal bands qualify, and these tend to produce front-loaded
hyetographs. The per-station quartile composition
(Fig.~\ref{fig:quartile_percent}) makes the same mechanism visible as a
continuous gradient rather than a binary dominance: the convective
biomes carry the largest first-quartile shares (46–51\%), and the
extratropical Mata Atlântica and Pampa redistribute roughly five
percentage points of that mass into the second and third quartiles,
exactly the signature of an increasing frontal and stratiform
contribution to mid-storm rainfall.

The systematic positive departure of Brazilian Q1 curves from the
Huff (1967) reference ($F_\mathrm{Brazil} > F_\mathrm{Illinois}$
for $\tau < 0.8$) reinforces the convective interpretation: Brazilian
storms not only peak in Q1 more frequently, but within Q1 they
accumulate rainfall faster than the Illinois reference. The Pampa
shows the smallest departure (MAE = 0.019), consistent with its more
Illinois-like climate.

\subsection{Comparison with prior studies}
\label{sec:comparison}

Placing our results alongside the international and Brazilian
literature requires comparing the same quantity (the event-level
quartile composition) rather than the station-level dominance figure
discussed above. Table~\ref{tab:comparison} assembles the reported
quartile results from the principal published derivations, normalised
where possible to the fraction of events in each quartile.

\begin{table}[H]
  \centering
  \caption{Comparison of reported Huff quartile results across the
    principal published derivations. Values are the fraction of
    \emph{events} in each quartile (Q1–Q4) where available; ``Q1+Q2''
    denotes a combined figure reported jointly. The station-level Q1
    dominance of the present study (94.4\% of stations) is a different
    metric and is not directly comparable to the event-level fractions
    in this table (see Section~\ref{sec:mechanism}). n/r = not reported
    as a separable value.}
  \label{tab:comparison}
  \setlength{\tabcolsep}{4pt}
  \footnotesize
  \begin{tabular}{lllrrrrl}
    \toprule
    Study & Region & Climate & Q1 & Q2 & Q3 & Q4 & Notes \\
    \midrule
    \citeA{Huff1967}        & Illinois, USA      & Temperate cont.\ &
      30 & 36 & 19 & 15 & Q2 most frequent \\
    \citeA{AzliRao2010}     & Pen.\ Malaysia     & Tropical maritime &
      \multicolumn{4}{c}{Q2 dominant (n/r)} & $\sim$5{,}800 storms \\
    \citeA{Liang2017}       & Guangzhou, China   & Subtropical &
      \multicolumn{2}{c}{Q1+Q2 = 84} & \multicolumn{2}{c}{n/r} &
      peak at 33\% of duration \\
    \citeA{SantaCatarina2021} & Santa Catarina, BR & Subtropical high &
      46.7 & $\sim$28 & $\sim$12 & $\sim$13 & Lages station \\
    \citeA{SantaCatarina2021} & Santa Catarina, BR & Subtropical high &
      44.7 & $\sim$28 & $\sim$12 & $\sim$13 & São Joaquim station \\
    \citeA{Florianopolis2021} & Florianópolis, BR & Coastal subtrop.\ &
      \multicolumn{4}{c}{Q1 (Type I) predominant} & single city \\
    \textbf{This study}     & \textbf{Brazil (national)} &
      \textbf{Mixed trop./subtrop.} &
      \textbf{43.6} & \textbf{24.3} & \textbf{19.0} & \textbf{13.1} &
      290{,}164 events \\
    \bottomrule
  \end{tabular}
\end{table}

At the event level, our national Q1 fraction of 43.6\% sits within the
range of previous work and is fully consistent with the only
comparable Brazilian study. \citeA{Huff1967} reported quartile
frequencies of 30\%, 36\%, 19\%, and 15\% for Q1–Q4 in Illinois, so
that Q1 and Q2 were jointly dominant and Q2 was in fact the single most
frequent class, a temperate-climate pattern in which organised and
frontal systems contribute a substantial share of back-loaded storms.
\citeA{SantaCatarina2021} derived Huff curves for two stations in the
mountain region of Santa Catarina (Lages and São Joaquim) and found Q1
the most frequent class at 46.7\% and 44.7\% of events respectively,
with the remaining events distributed across Q2 (around 28\%), Q4
(around 13\%), and Q3 (around 12\%); these values are very close to our
national event-level split and, like ours, place Q1 as the modal class
without an absolute majority. \citeA{Florianopolis2021} similarly
reported Q1 (Type I) as the predominant pattern for the coastal city
of Florianópolis.

The contrast with the tropical-Asian studies is instructive. In
Peninsular Malaysia, \citeA{AzliRao2010} found Q2 rather than Q1 to be
the most frequent class across roughly 5,800 storms, which they
attributed to the prevalence of multi-hour organised convection that
sustains peak intensities into the second quarter of storm time. For
Guangzhou, China, \citeA{Liang2017} reported that 84\% of storms fall
in Q1 or Q2 combined, with the normalised time of peak rainfall at
$33 \pm 5\%$ of storm duration, a front-loaded regime broadly
comparable to ours but reported as a combined Q1+Q2 figure rather than
a pure Q1 fraction. These differences reinforce the central message of
this study: the dominant temporal pattern is climate-specific, and the
Q1 prevalence we document for Brazil cannot be assumed a priori from
results obtained in other tropical or temperate regions.

Taken together, our study extends the isolated Brazilian site studies
to a national scale with two to three orders of magnitude more stations
and events, and is the first to show that the Q1 dominance previously
noted at individual southern-Brazilian sites is a continental rather
than local phenomenon.

\subsection{Quartile dependence of the fit to the Huff reference}
\label{sec:quartile_fit}

Mapping the goodness-of-fit metrics quartile by quartile
(Fig.~\ref{fig:map_mae}) exposes a systematic structure that the
aggregate Q1 view conceals, and that has a coherent physical
interpretation. Two effects act together. Geographically, agreement
with the Huff (1967) reference weakens monotonically from the
extratropical south toward the deep tropics in every quartile, mirroring
the biome gradient already noted for the Q1 curves: the storm
population of the Pampa and southern Mata Atlântica, where frontal and
mixed systems resemble the Illinois sampling conditions, is described
well by the reference, whereas the intense convective storms of
Amazônia and the Caatinga depart from it most strongly.

Superimposed on this geographic gradient is a consistent ordering among
the quartiles themselves. Across the national network the second
quartile is reproduced best (median $D_\mathrm{max}$ = 0.085) and the
fourth quartile worst (median $D_\mathrm{max}$ = 0.103), with Q1 and Q3
intermediate. The deterioration of the Q4 fit is concentrated in the
convective biomes and is pronounced: the median Q4 $D_\mathrm{max}$
reaches 0.175 in Amazônia, the single largest biome–quartile departure
in the dataset. We interpret this as a direct consequence of the same
convective mechanism that drives the national Q1 dominance. Where deep
afternoon convection prevails, genuinely back-loaded storms are both
uncommon and physically atypical: a Q4 classification in such a regime
tends to arise from compound or stratiform-trailing events whose mass
curve is poorly approximated by the smooth, strongly back-loaded Huff
Q4 reference. By contrast, in the extratropical south, where frontal
systems routinely generate late-peaking rainfall, the Q4 curves are
both better populated and closer to the reference. The second quartile,
representing a near-central and ubiquitous storm shape, is the most
transferable across climates. This quartile dependence reinforces the
central conclusion of the study: a single transplanted reference is
least adequate exactly where the local storm climatology is most
distinct, and the case for locally derived curves is therefore
strongest for the off-modal quartiles and the convective biomes.

\subsection{Practical implications for Brazilian design-storm practice}

The results have direct implications for urban drainage design,
stormwater management, and flood modelling in Brazil. The Huff (1967)
Illinois Q1 reference (widely used as a default in Brazilian practice)
under-predicts the cumulative fraction of storm depth reaching the
drainage system in the first quarter of storm time by approximately
four percentage points nationally (median $\tau = 0.5$: empirical
0.86 vs. reference 0.82). Although seemingly small in absolute terms,
the design-hydrograph experiment (Section~\ref{sec:hydro_results},
Fig.~\ref{fig:hydro}) shows that this difference translates into a
systematic underestimation of design peak discharge by the imported
curves of about 8\% nationally, and as much as 11\% in the convective
Cerrado, together with an advance of the time-to-peak of roughly half
an hour in the convective interior. Errors of this magnitude propagate
directly into the sizing of culverts, channels, and detention
structures, and they act in the unconservative direction — the
transplanted reference under-predicts the peak — so their correction is
of practical consequence for flood-risk infrastructure.

This magnitude of bias is not merely a methodological detail: it is
comparable to the peak-discharge uncertainty attributed to climate
change itself, a source of uncertainty the water-resources community
already treats as a first-order design concern. Climate-change
projections of peak streamflow commonly report increases on the order
of 10--30\% depending on region, emissions scenario, and return period
\cite{Maurer2018, Morsy2024}, consistent with global flood-risk
assessments that project comparably large shifts in flood frequency
and exposure under continued warming \cite{Hirabayashi2013}. Just as
ignoring climate non-stationarity is known to bias infrastructure
design against historical rainfall statistics \cite{Cheng2014}, our
results show that ignoring the temporal structure of rainfall — by
defaulting to an imported reference pattern — introduces a bias of
comparable magnitude, even when catchment, soil, and design depth are
held fixed. Storm-pattern selection should therefore be treated as a
source of design uncertainty on par with, not subordinate to, climate
change, and warrants the same scrutiny in Brazilian engineering
practice.

It is worth examining why the updated curves produce higher design peaks
at \emph{every} catchment in the experiment with no exception.
This consistency is not a numerical artefact but a direct consequence
of the central climatological finding: every Brazilian biome Q1
median curve lies \emph{pointwise above} the Huff (1967) Illinois Q1
reference across the full range of normalised storm time $\tau$
(Fig.~\ref{fig:national}), meaning that the local pattern delivers a
larger cumulative fraction of storm depth before any given moment
than the imported reference does.
Because all comparisons use the biome-level \emph{median} Q1 curve
against the Illinois \emph{median} Q1 curve, the ordering of the two
curves is fixed before the hydrological calculation begins; the
design-hydrograph experiment then translates that fixed ordering into
a peak-discharge difference whose magnitude depends on catchment
properties but whose sign is already determined.
Under individual station Q1 curves (which exhibit station-to-station
scatter around the biome median), a small fraction of catchments would
likely yield negligible or even slightly negative $\Delta Q_p$,
particularly in the southern Mata Atlântica and Pampa where the
local curves most closely approach the Illinois reference
(biome $D_\mathrm{max}$ = 0.042 and 0.062, respectively).
The engineering implication is that, at the biome- or state-median
level, the use of the Illinois reference in Brazil is consistently
unconservative.
Engineers should therefore apply the locally derived curves not only
where the curves differ most from the reference (Cerrado) but also in
the extratropical south, where the difference is smaller yet still
directionally unconservative.

It is also worth noting that the alternating-block method
(blocos alternados), widely used in Brazilian engineering practice as
a simple design-storm construction tool, produces centred hyetographs
whose cumulative mass curve resembles the Q2 or Q3 Huff patterns.
Because the Q2 and Q3 curves are less front-loaded than the Brazilian
Q1 reference, designs based on the alternating-block method tend to
underestimate design peak discharge even more severely than those
using the imported Huff (1967) Q1 reference, making the case for
locally derived Q1 curves all the more compelling.

Nevertheless, it is worth noting that the event-based design of
hydraulic infrastructures using the SCS-CN method is widely adopted
in Brazilian engineering practice. This fact underscores the relevance
of the present study and its experimental findings, as they elucidate
the importance and potential consequences of properly selecting
rainfall temporal distribution in design procedures.

The polynomial coefficient tables provided at biome, state, and
municipality levels (Supplementary data) offer a locally calibrated
alternative. State-level curves are available for all 27 Brazilian
administrative units and are recommended for engineering applications
where biome-level aggregation is too coarse. Municipality-level curves
are available for 694 municipalities but are considered reliable (here
defined as $n \geq 5$ stations) for only 19 of these; the remainder
should be treated as preliminary estimates.

Practitioners should note that the 50th-percentile (median) curves
are appropriate for mean-behaviour design, while the P10 and P90
envelopes (available in the supplementary long-form curve tables)
correspond roughly to the conservative and liberal bounds.

\subsection{Data quality and methodology considerations}

The most consequential methodological decision in this study is the
minimum storm depth of 12.7~mm (replicating Huff, 1967). Our IETD
sensitivity analysis confirms that the dominant findings are robust
across a wide range of IETD values, but the minimum depth threshold
was not varied: raising it would further restrict the event population
to the most intense storms and might shift the Q1 fraction upward, while
lowering it to 1~mm (a common default in sub-daily rainfall processing)
would include drizzle events that may not represent the heavy storms
relevant to design.

The mixed temporal resolution of the station network (51.8\% at
60~min) is an acknowledged limitation. While Q1 dominance is
consistent across all three resolution classes (97.8\% at 15~min,
92.8\% at 60~min), the 60-min stations cannot capture events shorter
than four hours, potentially missing the most intense convective cells
that would most strongly reinforce Q1 dominance.

\subsection{Limitations}
\label{sec:limitations}

Several limitations qualify the results presented here. The most
important concerns spatial coverage: station density is substantially
lower in the Amazon interior and the semi-arid Northeast than in the
densely instrumented south and southeast, so the curves for the
under-sampled regions, and particularly the Pantanal ($n = 5$
stations), should be regarded as indicative rather than definitive.
A second limitation arises from the mixed temporal resolution of the
network, in which 51.8\% of accepted stations record at 60-minute
intervals and therefore cannot resolve sub-hourly rainfall structure;
although the dominant-quartile classification is stable across
resolution classes, the curves at coarse-resolution stations
necessarily represent somewhat longer events. The analysis also
relies on a single data source (the ANA telemetric network) without
cross-validation against the INMET automatic stations, the CEMADEN
sub-hourly gauges, or radar precipitation estimates, each of which may
carry different sampling and missing-data characteristics.

Three further limitations concern the temporal and methodological
scope of the study. Because the derived curves represent the
2010–2025 climatological period, they assume stationarity and do not
capture any long-term trend in storm temporal structure, which may be
relevant for design horizons extending decades into the future
\cite{Ballarin2022}. A
single inter-event time definition of 6~h was applied uniformly across
the country; while the sensitivity analysis shows that the principal
findings are robust to this choice, region-specific IETD calibration
may nonetheless refine the curves in climatically distinct biomes.
The municipality-level products must be used with care, since
97.3\% of municipalities are represented by fewer than five accepted
stations and their curves consequently carry substantial uncertainty.

Finally, the design-hydrograph experiment is deliberately a screening
exercise rather than a calibrated flood study, and its scope is bounded
by the validity envelope of the SCS-CN method. It applies only to small,
infiltration-excess headwater catchments (5–250~km$^2$) in the Cerrado,
Caatinga, Mata Atlântica, and Pampa; the Amazônia and Pantanal biomes
were excluded a priori because their saturation-excess and wetland
runoff regimes lie outside the SCS-CN conceptual model, so the
experiment is not national in coverage. The reported changes are
relative comparisons in which the curve number, time of concentration,
and unit hydrograph are held identical between the two runs; this
isolates the temporal-pattern effect and causes systematic errors in
those inputs to cancel, but the absolute peak magnitudes are not
calibrated against observed hydrographs and should not be interpreted as
design values. The choice of time-of-concentration formula, curve-number
table, and unit-hydrograph shape would shift the absolute peaks but, as
relative-comparison factors, has limited influence on the reported
$\Delta Q_p$.

A subtler limitation relates to how the SCS-CN model handles
the temporal distribution of rainfall and runoff generation.
In the SCS-CN formulation, total storm runoff $Q$ is a function of
total rainfall depth $P$ and the curve number alone
($Q = (P - I_a)^2 / (P - I_a + S)$); it is independent of how
rainfall is distributed within the storm.
Consequently, the total runoff volume — and hence the runoff
coefficient $RC = Q/P$ — is identical for both the reference and the
locally derived Huff patterns, even though the hyetograph shapes differ
substantially.
The temporal pattern manifests only in the \emph{timing} of runoff
generation: a front-loaded Brazilian Q1 curve satisfies the initial
abstraction earlier, concentrating effective rainfall in the first
quarter of storm time and thereby driving the higher and earlier peaks
documented here.
This is a known conceptual constraint of the SCS-CN approach, which
lacks an infiltration-capacity mechanism linking rainfall intensity to
overland-flow generation.
In more physically-based rainfall–runoff frameworks — such as
Green–Ampt, Philip, or continuous soil-moisture accounting models — the
temporal distribution of rainfall is expected to influence total runoff
generation as well, because high early intensities can exceed the
soil's infiltration capacity even when the total depth would not
produce excess under a uniform distribution.
Whether the systematic front-loading of Brazilian storms relative to
the Illinois reference translates into additional runoff volume beyond
what the SCS-CN model captures is an open question that warrants
investigation with event-based continuous simulations or
infiltration-capacity models calibrated to Brazilian soils.

A further avenue for future research concerns the diurnal cycle
results documented here. Comparing the station-based diurnal cycle of
event initiation (Section~\ref{sec:diurnal}) against sub-daily
satellite remote sensing precipitation products — such as
IMERG \cite{Huffman2019} or GPM-IMERGHH — would provide an
independent, spatially continuous validation of the observed
afternoon convective maximum and its biome-to-biome variability,
contributing to the broader assessment of how well satellite products
capture storm temporal structure in South America.

\section{Conclusions}
\label{sec:conclusions}

This study presents the first national-scale derivation of empirical
Huff curves for Brazil, using 290,164 qualifying storm events
extracted from 1,045 ANA telemetric rain-gauge stations over the
period 2010–2025.

The central finding is that the first quartile is the dominant storm
pattern across the entire country: 94.4\% of stations are Q1-dominant,
reflecting Brazil's predominantly thermally driven convective rainfall
regime. This preponderance is remarkably robust, holding across all
six IBGE biomes, across the three data-collection timesteps present in
the network (15–60~min), and across a six-fold range of inter-event
time definitions (2–12~h). Crucially, however, the Brazilian Q1 curves
are systematically steeper than the Huff (1967) Illinois reference:
the national Q1 median curve exceeds the reference by up to
$\Delta F = 0.097$ in maximum absolute deviation, with the largest
positive departures occurring in the Amazônia and Cerrado biomes,
which are dominated by deep tropical convection, and the smallest in
the Pampa, whose extratropical climate most closely resembles the
Illinois context in which the original curves were calibrated.

We attribute this national Q1 prevalence to the rapid release of
thermodynamic instability by afternoon deep convection, which is the
dominant precipitation mechanism across most of Brazil and concentrates
rainfall in the first quarter of storm time. The interpretation is
supported by the seasonal signal, in which the austral winter (JJA)
dip in Q1 fraction to 33.6\% coincides with the increased frontal and
stratiform influence that characterises the subtropical south in that
season. The choice of a 6-h inter-event time definition is justified
on three independent grounds — replication of the original Huff (1967)
protocol, maximisation of the qualifying event count (290,164), and
the lowest curve-fit error relative to the reference among the tested
values — lending additional confidence to the derived curves.

Beyond these scientific findings, the study delivers a practical
resource for Brazilian hydrological engineering. Polynomial
coefficients and long-form percentile curves are provided at station,
biome, state, and municipality levels as open supplementary data and
through the \emph{Huff Curves BR Atlas} interactive web application
(Supplementary Material, Section~S8), offering a locally calibrated alternative
to the transplanted Illinois reference wherever Brazilian sub-daily
rainfall data are available. A controlled SCS-CN design-hydrograph
experiment confirms that the systematic upward departure of the
Brazilian curves from the Huff (1967) reference is not merely
statistical: across 579 small headwater catchments it raises the design
peak discharge by a median of 8\%, and by up to 11\% in the convective
Cerrado, with the imported curves erring in the unconservative
direction. Because this has direct consequences for the sizing of urban
drainage and flood-control infrastructure, we recommend that locally
derived curves replace the Illinois reference in Brazilian practice.

\section*{Supporting Information}

The Supporting Information describes the input datasets used in the
design-hydrograph experiment, the catchment-eligibility criteria, the
full mathematical formulation of the SCS-CN method, per-catchment
results, the per-station quartile distribution, and the
\emph{Huff Curves BR Atlas} interactive web application, which is
freely available at
\url{https://marcusnobrega-eng.github.io/Huff_Curves_BR}.

\section*{Open Research Section}

The ANA telemetric rainfall data are publicly available through the
ANA Hidroweb portal (\url{https://www.snirh.gov.br/hidroweb}).
All processed outputs — station-level, biome-level, state-level, and
municipality-level Huff curve coefficients, long-form percentile
curves, the design-hydrograph per-catchment results, and the Python
pipeline used to produce them — are archived on Zenodo
\cite{GomesJunior2026Zenodo}. The actively developed source code and
an interactive web atlas of the results are additionally available at
\url{https://github.com/marcusnobrega-eng/Huff_Curves_BR}.
The design-hydrograph experiment additionally uses publicly available
third-party datasets: HydroBASINS level-12 catchments
\cite{Lehner2013}, the Copernicus GLO-90 digital elevation model
\cite{CopernicusDEM}, SoilGrids 250~m soil-texture fractions
\cite{Poggio2021}, and the gridded Sherman-type intensity–duration–frequency surfaces of
\citeA{GomesJuniorIDF}.

\section*{Conflict of Interest Statement}

The authors declare that they have no known competing financial interests
or personal relationships that could have appeared to influence the
work reported in this paper.

\acknowledgments
The authors thank the Agência Nacional de Águas e Saneamento Básico
(ANA) for maintaining and openly distributing the telemetric
hydrometeorological network used in this study.

%

\end{document}